\documentclass[prd,preprint,amsmath,amssymb,aps,longbibliography]{revtex4-1}

\selectfont 

\usepackage{hyperref}
\usepackage{graphicx}       
\usepackage{dcolumn}        
\usepackage{bm}             
\usepackage{amssymb}
\usepackage{amstext}
\usepackage{tensor}

\newcommand{\nn}{\nonumber}

\newcommand{\gam}{\gamma}
\newcommand{\bet}{\beta}
\newcommand{\alp}{\alpha}

\newcommand{\sig}{\sigma}

\newcommand{\eps}{\epsilon}

\newcommand{\bnab}{\mathbf{\nabla}}

\newcommand{\beq}{\begin{equation}}
\newcommand{\eeq}{\end{equation}}

\newcommand{\Par}{\mathcal{P}}

\newcommand{\bxi}{\boldsymbol{\xi}}
\newcommand{\bM}{\mathbf{M}}
\newcommand{\bA}{\mathbf{A}}
\newcommand{\balp}{\boldsymbol{\alpha}}
\newcommand{\bbet}{\boldsymbol{\beta}}
\newcommand{\bgam}{\boldsymbol{\gamma}}

\newcommand{\opO}{\hat{\mathcal{O}}}
\newcommand{\opE}{\hat{\mathcal{E}}}

\begin{document}

 \title{Bound states of the Dirac equation on Kerr spacetime}

\author{Sam R.~Dolan}
\email{s.dolan@sheffield.ac.uk}
\affiliation{Consortium for Fundamental Physics, School of Mathematics and Statistics,
University of Sheffield, Hicks Building, Hounsfield Road, Sheffield S3 7RH, United Kingdom.}

\author{David Dempsey}
\email{ddempsey1@sheffield.ac.uk}
\affiliation{Consortium for Fundamental Physics, School of Mathematics and Statistics,
University of Sheffield, Hicks Building, Hounsfield Road, Sheffield S3 7RH, United Kingdom.}

\date{\today}

\begin{abstract}
We formulate the Dirac equation for a massive neutral spin-half particle on a rotating black hole spacetime, and we consider its (quasi)bound states: gravitationally-trapped modes which are regular across the future event horizon. These bound states decay with time, due to the absence of superradiance in the (single-particle) Dirac field. We introduce a practical method for computing the spectrum of energy levels and decay rates, and we compare our numerical results with known asymptotic results in the small-$M \mu$ and large-$M \mu$ regimes. By applying perturbation theory in a horizon-penetrating coordinate system, we compute the `fine structure' of the energy spectrum, and show good agreement with numerical results. We obtain data for a hyperfine splitting due to black hole rotation. We evolve generic initial data in the time domain, and show how Dirac bound states appear as spectral lines in the power spectra. In the rapidly-rotating regime, we find that the decay of low-frequency co-rotating modes is suppressed in the (bosonic) superradiant regime. We conclude with a discussion of physical implications and avenues for further work.
\end{abstract}

\maketitle

\section{Introduction}

Einstein's theory of general relativity (1915) and Dirac's relativistic wave equation (1928) are cornerstones of physics and `two households alike in dignity'. Although Dirac may have remarked that `it is more important to have beauty in one's equations than to have them fit experiment', these theories have also endured ever-more-stringent experimental tests. 

Thus far, there has been little opportunity to test the beautiful theory of fermions on curved spacetimes, developed by Schr\"odinger, Bargmann, Pauli \cite{Pauli:1933gc} (1933) and others \cite{Halpern:Heller:1935, Brill:1957fx}; except for in a restrictive  `Newtonian' setting \cite{Nesvizhevsky:2002ef, Nesvizhevsky:2003ww}. This may change in forthcoming decades. Black holes -- a radical consequence of Einstein's theory which appear to play a pivotal role in the development of galactic structure -- may act as crucibles for stringent tests of strong-field phenomena \cite{Berti:2015itd}, and the interaction of fermionic fields with black holes, the topic of this article, is of interest from a range of perspectives. 

In standard experience, there is a vast difference between the typical gravitational and quantum-mechanical length scales. A gravitating mass $M$ interacting with a quantum field of mass $\mu$ is characterized by a dimensionless parameter
\beq
 \frac{M \mu}{m_{\text{Pl}}^2}  = \frac{G M \mu}{\hbar c} \sim \frac{r_h}{\lambda_C} ,
\eeq
where $r_h$ is the horizon radius of the gravitating mass, $\lambda_C$ is the Compton wavelength of the field, and $m_{\text{Pl}}$ is the Planck mass. As $r_h$ is typically measured in kilometres, and $\lambda_C$ in pico or femtometres, it is clear that $M \mu \gg 1$ in standard astrophysical scenarios (henceforth we will typically omit dimensionful constants, setting $G = \hbar = c = m_{\text{Pl}} = 1$). Yet one may also envisage scenarios in which this is not the case, e.g., for (i) standard-model fields bound to `light' primordial black holes, and (ii) ultra-light fields (arising from e.g.~string theory compactifications) bound to galactic black holes \cite{Arvanitaki:2009fg, Arvanitaki:2014wva}. It has been suggested that ultra-light fields could provide a resolution of the dark matter problem (see e.g.~Refs.~\cite{Nucamendi:2000jw, Marsh:2010wq, Burt:2011pv}). 

The `no-hair' (Israel-Carter) conjecture \cite{Israel:1967za, Carter:1971zc, Ruffini:1971bza} suggests that, once perturbed, a black hole will rapidly return to a stationary state, changing only a small number of physical parameters in the process: its mass $M$, angular momentum $J$, and charge $Q$. The conjecture has been elevated to a theorem for various scenarios involving massless scalar, electromagnetic and gravitational fields \cite{Israel:1967za, Carter:1971zc, Ruffini:1971bza, Bekenstein:1995un, Bekenstein:1996pn, Chrusciel:2012}. However, if a perturbing field is endowed with mass, the quantum/classical correspondence principle suggests that the `no-hair' picture is incomplete. Outside the innermost stable circular orbit ($r_{\text{isco}} = 6GM / c^2$ for Schwarzschild black holes), a compact body of mass $\mu$ may orbit a black hole indefinitely, at least up to a gravitational radiation-reaction timescale $\tau_{\text{rad}} \sim (M/\mu) (GM/c^3)$. In the semi-classical regime $\lambda_c \ll r_\text{h}$ (i.e.~$M\mu \gg 1$) a straightforward WKB analysis (see Appendix \ref{sec:wkb}) suggests that a massive field possesses bound states $E < \mu c^2$ localized around orbiting timelike geodesics. For circular orbits in the weak-field regime ($r_0 \gg M$),
\beq
E  \approx  \mu c^2 - \frac{1}{2} \frac{GM\mu}{r_0} + \hbar \Omega_0 (n + \tfrac{1}{2}) , \quad \quad n \in \mathbb{N},  
\label{eq:intro-wkb}
\eeq
where the first term is the rest mass energy, the second term is the `Newtonian' binding energy, and the third term gives the energy-level spacing in terms of the orbital frequency $\Omega_0 = \sqrt{GM} / r_0^{3/2}$. In cases where  $\hbar \Omega_0 \ll \mu c^2$ the discrete spectrum becomes indistinguishable from a continuum. 

There is a subtlety in the above picture. Localized states are (generically) asymmetric under time-reversal, due to non-Hermiticity in the Dirac equation on black hole spacetimes. Flux is absorbed through the event horizon, which acts as a one-way membrane. Notwithstanding, lifetimes become exponentially-long when an angular momentum barrier separates the orbit from the event horizon. That is, we have $\tau \sim e^{M\mu \beta}$ where $\beta$ is some tunnelling coefficient which depends only on the ratio of $M\mu / (\ell + 1/2)$, with $\ell$ the orbital angular momentum number. 

How does a black hole's rotation affect this picture? Consider a Kerr black hole, with angular momentum $a M$. It is well-established that low-frequency modes of bosonic fields may be amplified by a process known as superradiance.  Localized bosonic states within the superradiant regime $\omega \tilde{\omega} < 0$, where $\tilde{\omega} = \omega - m \Omega_H$ with $\Omega_H = a / (2 M r_h)$ the angular frequency of the horizon, and $m$ the azimuthal number of the mode, may grow with time, causing a superradiant instability to develop: this is the so-called `black hole bomb' scenario \cite{Press:1972zz,Damour:1976kh}. Although growth, like decay, is exponentially-suppressed in the semi-classical regime $M\mu \gg 1$ (with e-folding time $\tau \gtrsim 10^7 \exp(1.84 M \mu)$ \cite{Zouros:1979iw}), it  becomes significant for $M \mu \sim O(1)$ (with $\tau_{\text{min}} \approx 5.88 \times 10^6 GM/c^3$ for the $\ell = m =1$ mode of the scalar field at $M \mu = 0.45$ and $a=0.997M$ \cite{Furuhashi:2004jk,Cardoso:2004nk,Dolan:2007mj,Dolan:2012yt}). By contrast, the (single-particle) Dirac equation is not subject to superradiance \cite{Martellini:1977qf, Iyer:1977cj, Iyer:1978du, Gueven:1977dq, Wagh:1986cz}, and thus all states decay. See Ref.~\cite{Brito:2015oca} for a recent summary of superradiance and associated phenomena, and Refs.~\cite{Brito:2014wla, Yoshino:2015nsa, Zilhao:2015tya} for recent work on the evolution of superradiant instabilities. 

Many authors \cite{Damour:1976kh,Zouros:1979iw,Detweiler:1980uk,Furuhashi:2004jk,Cardoso:2004nk,Dolan:2007mj,Rosa:2009ei,Ternov:1978gq, Ternov:1980st, Gal'tsov:1983, Ternov:1986bf, Gaina:1988dp, Ternov:1988xy, Gaina:1988nf, Gaina:1989bf, Lasenby:2002mc, Sturm:2007yf, Grain:2007gn, Hartman:2009qu, Laptev:2006, Pekeris:1989, Gair:thesis, Dokuchaev:2014} have considered the `bound states' of massive fields on Schwarzschild, Kerr and Kerr-Newman spacetimes. A range of terminology has been used, including `quasiresonances',  `quasilevels' \cite{Gaina:1992}, `quasistationary states', `quasibound states' \cite{Gal'tsov:1983}, `dynamical resonance states' \cite{Zhou:2013dra}, `wigs' \cite{Barranco:2012qs} and `graviatoms' \cite{Laptev:2006}; here we use `bound states' generically, without any implication of time-reversal symmetry. 
Bound states are not to be confused with `quasinormal modes', which are radiative in character. Loosely speaking, quasinormal modes are associated with a maximum (rather than a minimum) in the effective potential; they were studied in the massless Dirac case in Refs.~\cite{Cho:2003qe, Jing:2005dt, Jing:2005pk}. 

Regardless of differences in terminology, in the regime $M \mu / (\ell + 1/2) \lesssim 1$, a unified picture of bound states emerges. Let $\omega = \omega_R + i \omega_I$ denote the real and imaginary parts of the mode frequency, with $\omega_I < 0$ ($\omega_I > 0$) corresponding to decay (growth). To lowest order, the spectrum of spin-zero and spin-half particles is hydrogenic, with
\beq
\omega_R / \mu  \approx 1 - \frac{(M \mu)^2}{2 n^2} , \label{eq:omR-hydrogen}
\eeq
with $n \in 1, 2, \ldots$ the principle quantum number.  In this regime, the bound states have a typical radius of $ n a_0$, where $a_{0} = (M\mu)^{-2} (GM/c^2)$ is the gravitational `Bohr radius'. The imaginary part of a \emph{bosonic} field of spin $s$ (scalar $s=0$, Proca $s=1$ and massive graviton $s=2$) scales according to \cite{Detweiler:1980uk, Rosa:2011my, Witek:2012tr, Pani:2012vp, Pani:2012bp}
\beq
\omega_I / \mu \propto - \left(1 - \frac{m \Omega_H}{\mu}\right) (M \mu)^{4 l + 2 S + 5} , \label{eq:omI:boson}
\eeq
where $S \in \{ -s, -s+1, \ldots, s \}$ is the spin projection. By comparison, for the Dirac field it was shown 
in Ref.~\cite{Ternov:1980st} that, for $a \ll M$,
\beq
\omega_I / \mu \propto -(M \mu)^{4 l + 2 S + 4} ,  \label{eq:omI:fermion}
\eeq 
where $S \in \{ -1/2, 1/2 \}$ is the spin projection. This highlights a key point: in the regime $\mu < m \Omega_H$, bosonic modes become unstable, due to superradiant growth, whereas fermionic modes remain stable \cite{Martellini:1977qf, Iyer:1977cj}. A more subtle point here is that the index in Eq.~(\ref{eq:omI:fermion}) is smaller by $1$ than would be anticipated from Eq.~(\ref{eq:omI:boson}).

In this paper we formulate a practical method for computing Dirac bound states on the Kerr spacetime. Our aim is to use highly-accurate numerical data to verify and test various asymptotic results, such as those above; and to investigate the rich phenomenonology of the intermediate regime $M\mu \sim O(1)$. 

Rather remarkably, the massive Dirac equation on the Kerr spacetime admits a complete separation of variables. This reduces our problem to the analysis of (coupled) ordinary differential equations. Separability was first shown by Unruh \cite{Unruh:1973} in 1973 for the massless field, and by Chandasekhar \cite{Chandrasekhar:1976ap, Chandrasekhar:1983} in 1976 for the massive field. Whereas Unruh's analysis used the standard 4-spinor formalism, Chandrasekhar's analysis employed the Newman-Penrose 2-spinor formalism. In 1979, Carter \& McLenaghan \cite{Carter:1979fe} showed that the Dirac operator commutes with a `generalized total angular momentum operator' constructed from the Killing-Yano tensor. In 1984, Kamran \& McLenaghan \cite{Kamran:1984mb} described (see theorem 3) a class of spacetimes for which the Dirac equation admits a separation of variables in the Weyl representation. In 1993, McKellar \emph{et al.}~\cite{McKellar:1993ej} conducted an explicit separation of variables in the 4-spinor formalism, using the Pauli-Dirac representation. Finster and collaborators \cite{Finster:1999ry, Finster:2000jz, Finster:2001vn, Finster:2008bg}, and others \cite{Belgiorno:2008hk, Belgiorno:2008xn} have also used the 4-spinor formalism, along with the Weyl representation. 

From one point of view, superradiance is a necessary consequence of the second law of black hole thermodynamics: the horizon area $A$ is a non-decreasing function of time. How, then, is the absence of superradiance consistent with this law? A key assumption underpinning the second law is that the weak energy condition holds, i.e., $-T_{\mu \nu} t^\mu t^\nu \ge 0$ for any timelike vector field, $t_\mu t^\mu < 0$, where $T_{\mu \nu}$ is the stress-energy tensor. This condition is violated by the (single-particle) Dirac field \cite{Unruh:1973}.


%
%
%
%
%
%
%
%
%
%
%
%
%
%
%
%


This paper is organised as follows. In Sec.~\ref{sec:formalism} we review the formalism for describing a neutral spin-half field on a Kerr black hole spacetime. In Sec.~\ref{sec:methods} we describe our methods for calculating the bound state spectrum. In Sec.~\ref{sec:results} we explore the bound state spectrum on Schwarzschild and Kerr spacetimes, developing and testing asymptotic results in the regimes $M\mu \ll 1$ and $M\mu \gg 1$. In Sec.~\ref{sec:conclusions} we discuss our findings.

Throughout, we set $G = c = \hbar = 1$ and adopt a metric signature $+2$ (except for in Appendix \ref{appendix:fine-structure}). Coordinate indices are denoted with Greek letters $\alpha, \beta, \gamma \ldots$ and tetrad basis indices with Roman letters $a,b,c,\ldots$. Symmetrization and antisymmetrization of indices is denoted with round and square brackets, $()$ and $[]$, respectively. We use $\partial_\mu$, $\nabla_\mu$ and $\hat{D}_\mu$ to denote partial, covariant and spinor derivatives, respectively.

\section{Formalism\label{sec:formalism}}
In this section we outline the formalism for describing spin-half fields on the Kerr black hole spacetime (N.B.~expert readers may wish to proceed immediately to Sec.~\ref{sec:methods}). 

\subsection{Kerr spacetime}

\subsubsection{Coordinate systems}
The region outside the event horizon of a Kerr black hole may be described with the Boyer-Lindquist coordinate system $\{ t , r, \theta, \phi \}$, in which the line element $ds^2 = g_{\alp \bet} dx^\alp dx^\bet$ takes the form
\begin{eqnarray}
ds^2 = - \left(1 - \frac{2Mr}{\rho^2} \right) dt^2 - \frac{4 a M r \sin^2 \theta}{\rho^2} dt d\phi + \frac{\rho^2}{\Delta} dr^2 \nn \\
+ \rho^2 d\theta^2 + \left( r^2 + a^2 + \frac{2M r a^2 \sin^2 \theta}{\rho^2} \right) \sin^2 \theta d\phi^2 , 
\end{eqnarray}
where $\rho^2 = r^2 + a^2 \cos^2 \theta$ and $\Delta = r^2 - 2Mr + a^2$. The Kerr spacetime has two horizons at $r_\pm = M \pm \sqrt{M^2 - a^2}$, and two stationary limit surfaces, at $r_{S\pm} = M \pm \sqrt{M^2 - a^2 \cos^2 \theta}$. The angular velocity of the event horizon is $\Omega_H = a / 2Mr_+$.

A deficiency of the Boyer-Lindquist system is that it takes an infinite coordinate time $t$ for ingoing geodesics to reach the outer horizon at $r = r_+$. To properly describe such geodesics, we may employ the `ingoing Kerr' coordinate system $\{ \tilde{t}, \tilde{r}, \tilde{\theta}, \tilde{\phi} \}$ with $\tilde{r} = r$, $\tilde{\theta} = \theta$ and
\beq
d\tilde{t} = dt + \frac{2 M r}{\Delta} dr, \quad \quad d\tilde{\phi} = d\phi + \frac{a}{\Delta} dr .
\label{eq:coordtransform}
\eeq
Explicitly, $\tilde{t} = t + \alpha(r)$, $\tilde{\phi} = \phi + \beta(r)$ where
\begin{align}
\alpha(r) &= \frac{2Mr_+}{r_+ - r_-} \ln|r - r_+| - \frac{2Mr_-}{r_+ - r_-} \ln|r-r_-| ,  \label{eq:alpha} \\
\beta(r) &= \frac{a}{r_+ - r_-} \ln \left| \frac{r-r_+}{r-r_-} \right| . \label{eq:beta}
\end{align}
In this coordinate system the inverse metric takes a simple form,
\begin{equation}
\tilde{g}^{\mu \nu} = \frac{1}{\rho^2} \begin{pmatrix} - \left( \rho^2 + 2Mr \right) & 2Mr & 0 &0 \\ 2Mr & \Delta & 0 & a \\ 0 & 0 & 1 & 0 \\ 0 & a & 0 & \tfrac{1}{\sin^2 \theta} \end{pmatrix}
\end{equation}
and the ingoing principal null geodesics are simply given by $dr = -d\tilde{t}$. In either coordinate system, the metric determinant is given by $\sqrt{-g} = \rho^2 \sin\theta$.  

\subsubsection{Canonical tetrad}
Let us introduce a tetrad of vectors $e_a^\alpha = \{e_0^\alpha, e_1^\alpha, e_2^\alpha, e_3^\alpha \}$ which we take to be an orthonormal basis, i.e., $g_{\alpha \beta} e_a^\alpha e_b^\beta = \eta_{\alpha \beta}$ where $\eta_{ab} = \text{diag}(-1, 1, 1, 1)$. Roman and Greek letters are used for tetrad and coordinate indices, respectively. Roman indices are raised and lowered with $\eta_{ab}$. It follows that $e^a_\alpha = \eta^{a b} g_{\alpha \beta} e_b^\beta$ and $g_{\alp\bet} = \eta_{ab} e_{\alp}^a e_{\bet}^b$. 

We will employ the `canonical' orthonormal (symmetric) tetrad for the Kerr spacetime introduced by Carter \cite{Carter:1968ks, Wiltshire:2009zza}, viz.,
\begin{align}
e^t_0 &= \frac{r^2+a^2}{\rho \sqrt{\Delta}} , &  e^\phi_0 &= \frac{a}{\rho \sqrt{\Delta}} , \nn \\
e^r_1 &= \sqrt{\Delta} / \rho , & e^\theta_2 &= 1 / \rho , \nn \\
e^t_3 &= \frac{a \sin \theta}{\rho} , & e^\phi_3 &=  \frac{1}{\rho \sin \theta}  ,  \label{eq:canonical-inverse}
\end{align}
with inverse components given in (\ref{eq:canonical}) 
such that $ds^2 = \eta_{ab} (e_\mu^a dx^\mu) (e_\nu^b dx^\nu)$.
It is straightforward to find the components of the canonical tetrad in the ingoing Kerr coordinate system, using $\tilde{e}_j^\mu = \frac{\partial \tilde{x}^\mu}{\partial x^\nu} e_j^\nu$. 

\subsubsection{Spacetime symmetries and conservation laws}
The spacetime admits two independent Killing vectors, $k^\mu = (\partial_t)^\mu$ and $h^\mu = (\partial_\phi)^\mu$, with the defining properties $k_{(\mu ; \nu)} = 0 = h_{(\mu ; \nu)}  $. Furthermore, the spacetime admits a Killing-Yano tensor $f_{\mu \nu}$, with the defining properties $f_{\mu \nu} = f_{[\mu \nu]}$ and $f_{\mu \nu ; \sigma} = f_{[\mu \nu ; \sigma]}$, namely
\beq
f_{\mu \nu} = 2 \left( a \cos \theta \, e_{[\mu}^0 e_{\nu]}^1 + r \, e_{[\mu}^2 e_{\nu]}^3 \right) .
\eeq 

Now consider the stress-energy tensor $T_{\mu \nu}$ associated with the field, which is symmetric in its indices and divergence-free ($\nabla_\mu \tensor{T}{^{\mu \nu}} = 0$). With the Killing vectors, we may form two independent divergence-free vectors, $J_{(E)}^\mu = \tensor{T}{^\mu _\nu} k^\nu = \tensor{T}{^\mu _t}$ and $J_{(J)}^\mu = \tensor{T}{^\mu _\nu} h^\nu = \tensor{T}{^\mu _\phi}$, associated with energy and azimuthal angular momentum, respectively.  A third divergence-free vector $J^\mu_{\Psi}$ is given by the Dirac probability current (see Secs.~\ref{subsec:current} and \ref{subsec:dirac-current}). 

From each divergence-free vector ($\nabla_\mu J^\mu = 0$) one may obtain a conservation law via Gauss' theorem, 
\beq
\int_{\mathcal{V}} \nabla_\mu J^\mu d V = \int_{\partial \mathcal{V}} J^\mu d \Sigma_\mu,  
\eeq
where $\mathcal{V}$ is a contiguous four-volume bounded by a three-volume (hypersurface) $\partial \mathcal{V}$. Here, the volume element is $dV = \sqrt{-g} \, dt dr d\theta d\phi$, and the 3-surface element $d\Sigma_\mu$ is defined in terms of the metric induced on the boundary hypersurface in the standard way \cite{poisson2004relativist}. We may construct a four-volume of infinitessimal extent $\Delta t$ confined between twin spacelike hypersurfaces at $t = t_0$ and $t=t_0 + \Delta t$ and twin timelike hypersurfaces $r = r_1$ and $r = r_2$ (where $t_0$ and $r_2 > r_1 \ge r_+$ are constants). This construction leads to a quasi-local conservation law in the form
\beq
\frac{\partial}{\partial t} \left\{ \int_{r_1}^{r_2} \oint ( \rho^2 J^t ) \,  d\Omega dr \right\} =  -
\left[ \oint (\rho^2 J^r) d\Omega \right]_{r_1}^{r_2} . \label{eq:conservation1}
\eeq
where $d\Omega = \sin\theta d\theta d\phi$.
Similarly, by considering hypersurfaces of constant $\tilde{t}$, we may obtain the corresponding expression in the ingoing-Kerr coordinate system,
\beq
\frac{\partial}{\partial \tilde{t}} \left\{ \int_{r_1}^{r_2} \oint ( \rho^2 \tilde{J}^{\tilde{t}} )  d\tilde{\Omega} dr \right\} =  -
\left[ \oint (\rho^2 \tilde{J}^r) d\tilde{\Omega} \right]_{r_1}^{r_2} . \label{eq:conservation2}
\eeq
where $\tilde{J}^{\tilde{t}} = J^t + \frac{2Mr}{\Delta} J^r$ and $\tilde{J}^{r} = J^r$. Note that the $t = \text{const}$ and $\tilde{t} = \text{const}$ hypersurfaces are rather different in character, as the former approaches the bifurcation point, whereas the latter penetrates the future horizon. Physically, we expect (and find) that the ingoing-Kerr quantity $\tilde{J}^{\tilde{t}}$ is finite as $r \rightarrow r_+$, whereas the Boyer-Lindquist version $J^t$ is  not. 
 
\subsection{Spin-half fields}

\subsubsection{The Dirac equation\label{subsec:gamma-matrices}}
The Dirac equation on a curved spacetime (in signature $-+++$) takes the form \cite{Brill:1957fx}
\beq
\left( \gamma^\nu \hat{D}_\nu  - \mu \right) \Psi  = 0 , \label{eq:dirac}
\eeq
where $\mu$ is the field mass, $\Psi$ is a Dirac four-spinor, $\gamma^\nu$ are Dirac four-matrices, and the spinor covariant derivative $\hat{D}_\nu$ is
\beq
\hat{D}_\nu =  \partial_\nu - \Gamma_\nu .
\eeq
The spinor connection matrices $\Gamma_\nu$ are defined, up to an additive multiple of the unit matrix, by the relation
\beq
\partial_\nu \gamma^\mu + \tensor{\Gamma}{^\mu _{\nu \lambda}} \gamma^\lambda - \Gamma_\nu \gam^\mu + \gam^\mu \Gamma_\nu = 0. 
\label{eq:Gamdef}
\eeq
where $ \tensor{\Gamma}{^\mu _{\nu \lambda}}$ is the affine connection. A suitable choice satisfying (\ref{eq:Gamdef}) makes use of the spin connection $\omega_{\alpha \, b c}$ (described below), 
\beq
\Gamma_\alpha = -\frac{1}{4} \omega_{\alpha \, b c} \hat{\gamma}^b \hat{\gamma}^c .  \label{eq:Gam}
\eeq
Here, $\gamma^{\alpha}$ and $\hat{\gamma}^{a}$ denote sets of $4 \times 4$ matrices satisfying the anticommutation relations
\begin{align}
\{ \gamma^\alpha, \gamma^\beta \} &= 2 g^{\alpha \beta} I_4 ,   & 
 \{ \hat{\gamma}^a, \hat{\gamma}^b \}  &= 2 \eta^{ab} I_4, \label{gamanti}
\end{align}
where $\{A,B\} = A B + B A$. Note that the former set $\gamma^\alpha$ are functions of spacetime position, whereas the latter $\hat{\gamma}^a$ have constant components. The two sets may be related with any orthonormal tetrad,
\beq
\gamma^\alpha =  e_a^\alpha \hat{\gamma}^a .
\eeq
We will employ the canonical tetrad of Eq.~(\ref{eq:canonical-inverse}). Matrix indices are raised/lowered in the standard way: $\hat{\gamma}_a = \eta_{ab} \hat{\gamma}^b$ and $\gamma_\alpha = g_{\alp \bet} \gamma^\bet$.


\subsubsection{The spin connection}
The spin-connection $\omega_{\mu a b}$ arises naturally when one considers how a generalized covariant derivative $\nabla^{(\text{gen.})}_\mu$ should act upon tensors with mixed coordinate and basis components. We start with
\beq
\nabla^{(\text{gen.})}_\mu A^a_\alpha = \partial_\mu A^a_\alpha + \tensor{\omega}{_\mu ^a _b} A^b_\alpha - \tensor{\Gamma}{^\nu _{\alpha \mu}} A_\nu^a .
\eeq
Imposing metric compatibility ($\nabla_\gam^{(\text{gen.})} g_{\mu \nu} = 0$) leads to the usual definition for the affine connection $ \tensor{\Gamma}{^\nu _{\alpha \mu}} $. Imposing tetrad compatibility ($\nabla_\mu^{(\text{gen.})} e^a_\alpha = 0$) leads to
\beq
\tensor{\omega}{_\mu ^a _b} = e^a_\nu e^\lambda_b \tensor{\Gamma}{^\nu _{\mu \lambda}} - e_b^\lambda \partial_{\mu} e^a_{\lambda} .
\eeq
To obtain the spin connection without first calculating the affine connection one may use
\beq
\omega_{\mu a b} = \frac{1}{2} e^c_{\mu} \left(  \lambda_{abc} + \lambda_{cab} - \lambda_{bca} \right) ,
\eeq 
where $\lambda_{abc} = e_a^\mu \left(\partial_{\nu} e_{b \mu} - \partial_\mu e_{b\nu} \right) e_c^\nu$
 (see Ref.~\cite{Chandrasekhar:1983} for details). The spin connection for the canonical tetrad is listed in Appendix \ref{sec:spinconnection}.

\subsubsection{Current and stress-energy\label{subsec:current}}
The Dirac current is given by
\beq
J^\mu = \overline{\Psi} \gamma^\mu \Psi , \label{eq:current}
\eeq
where $\overline{\Psi} \equiv \Psi^\dagger \alpha$ is the Dirac conjugate, with $\Psi^\dagger$ denoting the usual Hermitian conjugate. The Hermitizing matrix $\alpha$ must satisfy the conditions
\beq
\alpha \gamma^\mu + {\gamma^\mu}^\dagger \alpha = 0, \quad \quad \partial_\mu \alpha + \Gamma_\mu^\dagger \alpha + \alpha \Gamma_\mu  = 0.
\eeq
We choose $\alpha = -\hat{\gamma}^0$ \cite{Casals:2012es}. 

The (symmetric) stress-energy $T_{\mu \nu}$ is given by
\beq
T_{\mu \nu} = \frac{i}{2} \left[ \overline{\Psi} \gam_{(\mu} \hat{D}_{\nu)} \Psi -  \left\{ \hat{D}_{(\mu} \overline{\Psi} \right\} \, \gam_{\nu)} \Psi \right] ,  \label{eq:stress-energy}
\eeq
where the covariant derivative of the conjugate spinor is $\hat{D}_{\mu} \overline{\Psi} = \partial_\mu \overline{\Psi} + \overline{\Psi} \Gamma_\mu$. The Dirac current and stress-energy are covariantly conserved, $\nabla_\mu J^\mu = 0 = \nabla_\mu T^{\mu \nu}$. 

\subsubsection{Matrix representation}
We will use the Weyl/chiral representation, in which
\beq
\tilde{\gam}^0 = \begin{pmatrix} O & I \\ I & O \end{pmatrix}, \quad \tilde{\gam}^i = \begin{pmatrix} O & \sig_i \\ -\sig_i & O \end{pmatrix} , \quad \quad i = 1,2,3,
\eeq
where $I$ is the $2\times2$ identity, $O$ is the $2 \times 2$ zero matrix, and $\sig_i$ are the Pauli matrices 
\beq 
\sig_1 = \begin{pmatrix} 0 & 1 \\ 1 & 0 \end{pmatrix}, \quad \sig_2 = \begin{pmatrix} 0 & -i \\ i & 0 \end{pmatrix}, \quad \sig_3 = \begin{pmatrix} 1 & 0 \\ 0 & -1 \end{pmatrix} ,
\eeq
which satisfy $\tfrac{1}{2} \{ \sig_i, \sig_j \} = \delta_{ij} I + i \epsilon_{ijk} \sig_k$. 

In fact, we will choose our matrices $\hat{\gamma}^i$ (Sec.~\ref{subsec:gamma-matrices}) to be a cyclic permutation of the set $\tilde{\gamma}^i$ multiplied by the unit imaginary, as follows: $\hat{\gamma}^1 = i \tilde{\gamma}^3$,  $\hat{\gamma}^2 = i \tilde{\gamma}^1$,  $\hat{\gamma}^3 = i \tilde{\gamma}^2$ and $\hat{\gamma}^0 = i \tilde{\gam}^0$. 
 Note that making the permutation is equivalent to relabelling the tetrad legs; 
alternatively one may take the view that we are applying a (constant) similarity transform $\gamma^\mu \rightarrow S \gamma^\mu S^{-1}$, $\Psi \rightarrow S \Psi$. Including the factor of $i$ is necessary in order to satisfy anticommutation relations (\ref{gamanti}) on a spacetime of positive signature. 

We will write the Dirac four-spinor as $\Psi = \begin{pmatrix} \psi_- \\ \psi_+ \end{pmatrix}$ where $\psi_+$ and $\psi_-$ are (left- and right-handed) two-spinors, which may be projected out from $\Psi$ with the operators $P_\pm = \frac{1}{2} \left(I \pm \hat{\gamma}_5 \right)$ where
\beq
\hat{\gam}^5 = i \hat{\gam}^0 \hat{\gam}^1 \hat{\gam}^2 \hat{\gam}^3 = \begin{pmatrix} -I & O \\ O & I \end{pmatrix} .
\eeq

\subsubsection{Separation of variables}
Let us introduce the complex quantity $\varrho = r + i a \cos \theta$, and its conjugate $\varrho^\ast = r - i a \cos \theta$, such that $\rho^2 = \varrho \varrho^\ast$. A short calculation (using Eq.~(\ref{eq:Gam}) and Appendix \ref{sec:spinconnection}) shows that the spin-connection matrices $\Gamma_\mu$ take the form given in Eq.~(\ref{eq:Gamma_mu}). Next we obtain 
\begin{align}
- \gam^\mu \Gamma_\mu &= \frac{i}{2} \begin{pmatrix} O & s^\ast_\theta \sig_1 + s^\ast_r \sig_3  \\ -(s_\theta \sig_1 + s_r \sig_3) & O \end{pmatrix} ,
\end{align}
where
$
s_r 
= e^r_1 \, \frac{1}{\varrho \sqrt{\Delta} } \frac{\partial}{\partial r} \left( \varrho \sqrt{\Delta} \right)$
and
$
s_\theta 
= e^\theta_2 \, \frac{1}{\varrho \sin \theta} \frac{\partial}{\partial\theta} \left( \varrho \sin \theta \right) 
$. This result for the spin-connection matrix suggests a natural ansatz for the wavefunction,
\beq
\Psi = \Delta^{-1/4} \begin{pmatrix} \varrho^{-1/2} \, \eta_- \\ {\varrho^\ast}^{-1/2} \eta_+ \end{pmatrix}
\label{eq:psi-ansatz1}
\eeq
where $\eta_\pm$ are two-spinors. 
%
Multiplying the Dirac equation (\ref{eq:dirac}) by $-i \Delta^{1/4} \rho \begin{pmatrix} {\varrho^\ast}^{1/2} & O \\ O & -\varrho^{1/2} \end{pmatrix}$ yields a pair of two-spinor equations,
\begin{align}
\left\{ 
\pm \frac{1}{\sqrt{\Delta}} \left[ (r^2+a^2)\partial_t + a \partial_\phi \right] + \sig_3 \sqrt{\Delta} \partial_r + \sig_1 \left(\partial_\theta + \tfrac{1}{2} \cot \theta\right) + \sig_2 \left(a \sin \theta \partial_t + \csc \theta \partial_\phi \right)
\right\} 
 \eta_\pm  & \nn \\ 
 \mp i \mu \left(r \mp ia \cos \theta\right) \eta_{\mp}  &= 0 .
\end{align}
Now, separating harmonic temporal and azimuthal dependence with
\beq
\eta_\pm(t,r,\theta,\phi) = e^{i (m \phi - \omega t)} \eta_\pm(r,\theta), 
\label{eq:psi-ansatz2}
\eeq
where $m$ is a half-integer, 
leads to a remarkable separation of the $r$ and $\theta$ parts, 
\begin{align}
\left\{ 
\mp \frac{i}{\sqrt{\Delta}} \left[ -(r^2+a^2)\omega + a m \right] + \sig_3 \sqrt{\Delta} \partial_r \right\} \eta_\mp
\mp i \mu r \eta_\pm & \nn \\
+ \left\{ \sig_1 \left(\partial_\theta + \tfrac{1}{2} \cot \theta\right) + i \sig_2 \left(-a \omega \sin \theta + m \csc \theta \right) \right\} \eta_\mp 
 +  a \mu \cos \theta \, \eta_{\pm} &=  0 , \label{eq:etapm}
\end{align}
(cf.~Eq.~(8) in Ref.~\cite{McKellar:1993ej}; see also Appendix \ref{sec:finster}). 
Introducing the ansatz
\begin{align}
\eta_+ &= \phantom{-} \begin{pmatrix} R_1(r) S_1(\theta) \\ R_2(r) S_2(\theta) \end{pmatrix} , &
\eta_- &= - \begin{pmatrix} R_2(r) S_1(\theta) \\ R_1(r) S_2(\theta) \end{pmatrix} ,
\label{eq:psi-ansatz3}
\end{align}
and multiplying by $\sig_3$ leads to twin pairs of coupled first-order ordinary differential equations,
\begin{align}
\sqrt{\Delta} \left( \partial_r - i K / \Delta \right) R_1 &= \left(\lambda + i \mu r \right) R_2 ,  \label{eq:R1} \\
\sqrt{\Delta} \left( \partial_r + i K / \Delta \right) R_2 &= \left(\lambda - i \mu r \right) R_1 ,  \label{eq:R2}
\end{align}
and
\begin{align}
\left( \partial_\theta + \frac12 \cot \theta - m \csc \theta + a \omega \sin \theta \right) S_1 &= \left(+ \lambda + a \mu \cos \theta \right) S_2 ,  \label{eq:S1} \\
\left( \partial_\theta + \frac12 \cot \theta + m \csc \theta - a \omega \sin \theta \right) S_2 &= \left(- \lambda + a \mu \cos \theta \right) S_1 , \label{eq:S2}
\end{align}
where $K = (r^2+a^2)\omega - a m$ and $\lambda$ is the separation constant. Eqs.~(\ref{eq:R1})--(\ref{eq:S2}) are equivalent to the coupled ordinary differential equations originally obtained by Chandrasekhar \cite{Chandrasekhar:1976ap, Chandrasekhar:1983}.

\subsubsection{Angular solutions}
In the non-rotating case ($a = 0$), the eigenvalues of the angular equations (\ref{eq:S1})--(\ref{eq:S2}) are integers $\lambda \in \mathbb{Z} \backslash 0 = \left\{ \ldots, -2, -1, +1, 2 \ldots \right\}$, and the solutions are  spin-weighted spherical harmonics (see Sec.~3 in Ref.~\cite{Dolan:2009kj}). The total angular momentum, $j$, a half-integer, is related to the $a=0$ eigenvalue in a simple way: $\lambda = \Par (j + 1/2)$, where $\Par = \pm 1$. We may define the orbital angular momentum, $\ell$, an integer, via $\ell = j + \tfrac{1}{2} \Par$. 

In the rotating case the solutions of Eqs.~(\ref{eq:S1})--(\ref{eq:S2}) are known as mass-dependent spheroidal harmonics (see Ref.~\cite{Dolan:2009kj} for a review). We let $S_1 = {}_-S_\Lambda$ and $S_2 = {}_+S_\Lambda$, where $\Lambda = \{j,m,\Par,a\omega,a\mu\}$ and we note the key symmetry relation
\beq
{}_\pm S_{\Lambda}(\theta) = \Par (-1)^{j+m} {}_\mp S_\Lambda(\pi - \theta),
\eeq
and normalization condition
\beq
\int_0^\pi \sin \theta \left(  \left|{}_{+}S_{\Lambda} \right|^2 +  \left|{}_{-}S_{\Lambda} \right|^2 \right) d \theta = \frac{1}{2\pi} .  \label{eq:Snorm}
\eeq

\subsubsection{Radial solutions}
Let $\mathbf{R} = \begin{pmatrix} R_1 \\  R_2 \end{pmatrix}$. In the near-horizon region $r \rightarrow r_+$, the `horizon-ingoing' radial solution has the asymptotic form
\begin{align}
\mathbf{R}_{\text{hor}}
\sim 
 \begin{pmatrix} \beta \sqrt{\Delta} \\ 1 \end{pmatrix}
\exp\left( - i \tilde{\omega} r_\ast \right) \label{eq:Rhor}
\end{align}
where
\begin{align}
\tilde{\omega} &= \omega - m \Omega_H , \\
\beta & = \frac{\lambda + i \mu r_+}{r_+ - r_- - 4iM \tilde{\omega} r_+} .
\end{align}
Here $r_\ast$ is the tortoise coordinate defined by
\beq
\frac{d r_\ast}{dr} = \frac{(r^2 + a^2)}{\Delta} . 
\eeq
The `horizon-outgoing' solution is obtained by the interchange $R_1 \rightarrow R_2^\ast$ and $R_2 \rightarrow R_1^\ast$. 

In the far-field region, $r \rightarrow \infty$, the propagating solutions take the form
\beq
\mathbf{R}^{\pm}_{\infty}
\sim
 \frac{r^{\pm \gamma} e^{\pm ipr}}{\sqrt{2 p (\omega \mp p)}} \left[ \begin{pmatrix} -\mu \\ \omega \mp p \end{pmatrix} + O(r^{-1}) \right] , \quad \quad \gamma = \frac{i M(2 \omega^2 - \mu^2)}{p} ,
\eeq
where $p = \sqrt{\omega^2 - \mu^2}$. The bound solutions ($\omega < \mu$) may be obtained by replacing $ip$ with $q = -\sqrt{\mu^2 - \omega^2}$ in the above. 

Multiplying (\ref{eq:R1}) and (\ref{eq:R2}) by $R_1^\ast$ and $R_2^\ast$, respectively, and taking the difference, leads to the Wronskian relationship
$\frac{d}{dr} \left( |R_1|^2 - |R_2|^2 \right) = 0$. Now let us consider a modal solution of the form
\begin{eqnarray}
A_h \mathbf{R}_{\text{hor}} = A_\infty^+ \mathbf{R}^{+}_{\infty} + A_\infty^- \mathbf{R}^{-}_{\infty} ,
\end{eqnarray}
where $A_h$ and $A_{\infty}^{\pm}$ are complex constants. From the constancy of the Wronskian, it follows immediately that
\beq
- \left|A_h\right|^2 = \left|A^+_\infty\right|^2 - \left|A^-_\infty\right|^2 ,  
\eeq
that is, $\mathcal{R} = 1 - \mathcal{T}$, where the reflection and transmission coefficients are $\mathcal{R} \equiv  \left|A^+_\infty\right|^2 /  \left|A^-_\infty\right|^2$ and 
 $\mathcal{T} \equiv  \left|A_h \right|^2 /  \left|A^-_\infty\right|^2$. As $\mathcal{T} \ge 0$, it is clear that superradiance does not occur in the modal reflection/transmission problem. 

\subsubsection{The Dirac current and absence of superradiance\label{subsec:dirac-current}}
We now insert (\ref{eq:psi-ansatz3}) (see also (\ref{eq:psi-ansatz1}) and (\ref{eq:psi-ansatz2})) into the definition of the Dirac current (\ref{eq:current}). The radial component is given by
\beq
J^r = \frac{1}{\rho^2} \left(\left|R_1\right|^2 - \left|R_2\right|^2 \right) \left(\left|S_1\right|^2 + \left|S_2\right|^2 \right).
\eeq
Let us refer back to the conservation law (\ref{eq:conservation1}). The integral of the radial current evaluated at the lower limit $r_1 = r_+$ has a natural interpretation as the flux passing into the horizon; we have 
\beq
\frac{d N}{d t} = \left. \left(\left|R_1\right|^2 - \left|R_2\right|^2 \right) \right|_{r=r_+}
\eeq
where $N$ is the number density. We saw in the previous section that the right-hand side is negative for horizon-ingoing solutions; hence $dN/dt \leq 0$. This provides further confirmation that superradiance is absent \cite{Unruh:1973, Martellini:1977qf, Iyer:1977cj, Iyer:1978du}.

Next, we obtain expressions for the components of the Dirac current in the canonical basis,
\begin{align}
J^{(0)} \equiv e^0_\mu J^\mu &= \frac{1}{\rho \sqrt{\Delta}} \left( \left|R_1\right|^2 + \left|R_2\right|^2 \right) \left( \left|S_1\right|^2 + \left|S_2\right|^2 \right) , \\
J^{(3)} \equiv e^3_\mu J^\mu &= \frac{1}{\rho \sqrt{\Delta}} 4 \text{Im}(R_1^\ast R_2) \, \text{Re}(S_1^\ast S_2) ,
\end{align}
The temporal component of the current, $J^t = e^t_0 J^{(0)} + e^t_3 J^{(3)}$, is
\beq
J^t = \frac{r^2+a^2}{\rho^2 \Delta} \left[\left(|R_1|^2 + |R_2|^2 \right)\left(|S_1|^2 + |S_2|^2 \right) + \frac{4a\sqrt{\Delta}\sin\theta}{r^2+a^2} \text{Im}(R_1^\ast R_2) \, \text{Re}(S_1^\ast S_2) \right] .  \label{eq:Jt-bl}
\eeq
Note that the factor of $\Delta$ in the denominator implies that $J^t$ diverges in the limit $r \rightarrow r_+$. This is due to the coordinate singularity in the Boyer-Lindquist system. We may instead consider the temporal component in the ingoing-Kerr coordinate system, given by $\tilde{J}^{\tilde{t}} =J^t + \frac{2 M r}{\Delta} J^r$ evaluated at $t = \tilde{t} - \alpha(r)$, $\phi = \tilde{\phi} - \beta(r)$ (cf.~Eq.~(\ref{eq:alpha})--(\ref{eq:beta})), i.e.~
\begin{align}
\tilde{J}^{\tilde{t}} &= \frac{1}{\rho^2} \left[\left(|\tilde{R}_2|^2 + \left(r^2 + 2Mr + a^2 \right) |\Delta^{-1/2}\tilde{R}_1|^2 \right)\left(|S_1|^2 + |S_2|^2 \right) \right. \nn \\
 & \quad \quad\quad \quad \left. + 4a\sin\theta \, \text{Im}(\Delta^{-1/2} \tilde{R}_1^\ast \tilde{R}_2) \, \text{Re}(S_1^\ast S_2) \right] , \label{eq:Jt-ingoing}
\end{align}
where
\beq
\tilde{R}_k(r) = e^{i\omega \alpha(r)} e^{-i m \beta(r)} R_k(r) , \quad \quad\quad k = 1, 2.
\eeq
From the asymptotic form of the ingoing solution $\mathbf{R}_{\text{hor}}$, Eq.~\ref{eq:Rhor}), it is clear that $\Delta^{-1/2}R_1 \sim O(1)$ and thus $\tilde{J}^{\tilde{t}}$ is \emph{finite} on the (future) horizon $r = r_+$, as expected.

\subsubsection{Violation of weak energy principle}
The weak energy condition states that $-T_{\mu \nu} t^\mu t^\nu \ge 0$ for any timelike vector $t^\mu$. Let us now introduce the quantity $\Xi = -T_{\mu \nu} e_0^\mu e_0^\nu$, noting that $e_0^\mu$ is indeed timelike. A short calculation gives $\Xi = \Xi_0 + \Xi_1$, where
\begin{align}
\Xi_{0} &= \frac{1}{\rho^2 \Delta} \left[ (r^2+a^2) \omega - a m \right] \left( |R_1|^2 + |R_2|^2 \right) \left( |S_1|^2 + |S_2|^2 \right)  \\
\Xi_{1} &= \frac{a}{2 \rho^4} \left[  \cos \theta  \left( |R_1|^2 + |R_2|^2 \right) \left( |S_1|^2 - |S_2|^2 \right) \right. \nn \\
& \quad \quad \quad \quad \quad  \left. -  \sin \theta  \frac{4r}{\sqrt{\Delta}} \, \text{Re}(R_1^\ast R_2) \, \text{Re}(S_1^\ast S_2)  \right] .
\end{align}
The first term $\Xi_0$ arises from the partial derivatives in Eq.~(\ref{eq:stress-energy}), and the second term $\Xi_1$ arises from the spin-connection matrices. The first term dominates over the second as $\Delta \rightarrow 0$. It is clear that $\Xi_0$ is not positive-definite; it is negative in the near-horizon region if $\omega < m \Omega_H$. Thus the weak-energy condition is violated for the Dirac equation on Kerr spacetime in the superradiant regime, at least at the `one particle' level (for second quantization, see Ref.~\cite{Unruh:1974bw}).



\section{Methods\label{sec:methods}}
In this section we outline practical frequency-domain (Sec.~\ref{subsec:method-freq}) and time-domain (Sec.~\ref{subsec:td-method}) methods for investigating the bound state spectrum. 

\subsection{Frequency-domain method\label{subsec:method-freq}}

We now formulate a practical method for computing the discrete spectrum of bound states. In essence, we wish to solve the radial equations (\ref{eq:R1})--(\ref{eq:R2}) subject to certain boundary conditions.  On the horizon, the solution should be finite in the Kerr-ingoing coordinate system; towards infinity, the solution should decay exponentially. Thus, the integral of the probability density (\ref{eq:Jt-ingoing}) should be finite across the exterior region $r \ge r_+$.

A powerful method for computing the spectrum of quasi-normal modes was introduced in Ref.~\cite{Leaver:1985ax}, in the context of massless fields. With an ansatz adapted to the boundary conditions, the differential equation for massless bosonic fields generates a three-term recurrence relation. The minimal solution of the recurrence relation is found by solving a continued-fraction equation. This approach was adapted to compute the bound state spectrum in Ref.~\cite{Dolan:2007mj}. We follow this approach below.

\subsubsection{Angular solutions}

Three-term recurrence relations for the angular equations (\ref{eq:S1})--(\ref{eq:S2}) were previously obtained in Refs.~\cite{Suffern, Kalnins:1992bf, Dolan:2009kj}. We follow the method of Ref.~\cite{Dolan:2009kj}, in which the mass-dependent spin-half spheroidal harmonics are decomposed in a basis of spin-half spherical harmonics, leading to a three-term recurrence relation for the expansion coefficients $\{b_k\}$ (where $k$ is a half-integer):
\begin{align}
\alp_k b_{k+1} + \bet_k b_k &= 0, \quad \quad k = |m|,  \\
\alp_k b_{k+1} + \bet_k b_k + \gam_k b_{k-1} &= 0, \quad \quad k = |m| + 1, \, |m|+2, \ldots 
\end{align}
where 
\begin{align}
\alp_k &= (a \mu + \epsilon_k a \omega) \frac{\sqrt{(k+1)^2 - m^2}}{2(k+1)} , \\
\bet_k &= \epsilon_k (k + 1/2) \left(1 - \frac{a m \omega}{k(k+1)} \right) + \frac{a \mu m}{2k(k+1)} - \lambda, \\
\gam_k &= (a \mu - \epsilon_k a \omega) \frac{\sqrt{k^2 - m^2}}{2k} ,
\end{align}
and $\eps_k = (-1)^{j-k} \mathcal{P}$ with $j = \ell \mp \mathcal{P} / 2$. The boundary conditions at the poles are  satisfied by the minimal solution of the recurrence relation, which is found by solving a continued-fraction equation for $\lambda$,
\beq
\bet_{|m|} - \frac{\alp_{|m|} \bet_{|m|+1}}{\bet_{|m|+1}-}\frac{\alp_{|m|+1} \gam_{|m|+2}}{\bet_{|m|+2}-} \ldots 
= 0 , \label{eq:ctdfrac}
\eeq
or one of its inversions.

\subsubsection{Radial solutions}
Let $R_2 = \sqrt{\Delta} R_+$ and $R_1 = R_-$, so that the radial equations (\ref{eq:R1})--(\ref{eq:R2}) become
\begin{align}
\left( \frac{d}{dr} - \frac{i K}{ \Delta}    \right)R_{-}     & =  ( \lambda +i \mu r ) R_{+}    ,     \\
\left( \frac{d}{dr} + \frac{i K -M +r }{ \Delta}    \right)    R_{+}    &=  \frac{ \lambda - i \mu r}{\Delta} R_{-} .
\end{align}
Near the horizon the `ingoing' solution goes as,
\begin{align}
\lim_{r \to r_+} R_+ &\to (r-r_+)^{\sig-1}, &  
\lim_{r \to r_+} R_- &\to (r-r_+)^{\sig},
\end{align}
and at infinity the decaying solution resembles
\begin{align}
\lim_{r \to \infty} R_+ &\to  r^{\nu-1} e^{qr} ,   &  
\lim_{r \to \infty} R_- &\to r^\nu e^{qr},
\end{align}
where
\begin{align}
\sig &=  \frac{1}{2} - \frac{2M i \tilde{\omega} r_+}{r_+ - r_-} , \\
\nu &= M\frac{\mu^2 -2\omega^2}{q},		\\
q &= -\sqrt{\mu^2 -\omega^2}.
\end{align}
We may write the solution in terms of a series,
\begin{align}
\left(
\begin{array}{c}
R_-\\
(r - r_+) R_+\\
\end{array}
\right) = 
(r-r_+)^{\sig} (r-r_-)^{-\sig+\nu} e^{qr} 
\sum \limits_{k=0}^\infty \boldsymbol{\xi}_k \left(\frac{r-r_+}{r-r_-}\right)^k  \label{eq:ansatz} .
\end{align}
Inserting Eq.~(\ref{eq:ansatz}) into the radial equations leads to a three-term \emph{matrix-valued} recurrence relation,
\begin{align}
\boldsymbol{\alpha}_0  \boldsymbol{\xi}_{1} +
\boldsymbol{\beta}_0  \boldsymbol{\xi}_{0} &=0, \\
\boldsymbol{\alpha}_k  \boldsymbol{\xi}_{k+1} +
\boldsymbol{\beta}_k  \boldsymbol{\xi}_{k}+
\boldsymbol{\gamma}_k  \boldsymbol{\xi}_{k-1} &=0,  \qquad k>0.
\end{align}
Here, the matrices are of the form
\begin{align}
\balp_k =  \left(
\begin{array}{cc}
\alpha_{k1} & \alpha_{k2} \\
0 & \alpha_{k4} \\
\end{array}
\right), \quad
\bbet_k =  \left(
\begin{array}{cc}
\beta_{k1} & \beta_{k2} \\
\beta_{k3} & \beta_{k4} \\
\end{array}
\right), \quad
\bgam_k =  \left(
\begin{array}{cc}
\gamma_{k1} & 0 \\
\gamma_{k3} & \gamma_{k4} \\
\end{array}
\right),
\end{align}
and the matrix coefficients are
\begin{align}
\alpha_{k1} &=  (k+\sigma+1)  (r_+ - r_-)   -   i ( 2 M \omega r_+ - a m ) ,  \\
\alpha_{k2} &=   -(\lambda + i \mu  r_+) (r_+-r_-) , \\
\alpha_{k4} &=  (k+\sigma+\tfrac{1}{2})  (r_+ - r_-)    +   i ( 2 M \omega r_+ - a m ) ,
\end{align}
\begin{align}
\beta_{k1} &=  (r_+-r_-) \left (q (r_+ - r_-) - 2 (k + \sigma)+ \nu \right)  + 2 i \left(2 a^2 \omega - a m \right) , \\
\beta_{k2} &= (\lambda +i \mu  r_-) (r_+-r_-) , \\
\beta_{k3} &= -(\lambda - i \mu  r_+) (r_+-r_-) ,  \\
\beta_{k4} &=  (r_+-r_-) \left (q (r_+ - r_-) - 2 (k + \sigma)+ \nu + 1\right)  - 2 i \left(2 a^2 \omega - a m \right) ,
\end{align}
\begin{align}
\gamma_{k1} &=  (k + \sigma - 1  - \nu) (r_+-r_-) - i ( 2 M \omega r_- - a m ) , \\
\gamma_{k3} &=   ( \lambda-i \mu r_-) (r_+-r_-) , \\
\gamma_{k4} &=  (k + \sigma - \tfrac{3}{2} - \nu)(r_+-r_-)  + i( 2 M \omega r_-  - a m) .
\end{align}

Matrix-valued three-term recurrence relations can be solved using matrix-valued continued fractions, as described in Refs.~\cite{Simmendinger:1999, Rosa:2011my}. We seek the roots of $\bM \bxi_0 = \mathbf{0}$, where
\beq
\bM \equiv   \bbet_0  -  \balp_0 \left[ \bbet_1 - \balp_1\left( \bbet_2 + \balp_2 \bA_2 \right) \bgam_2  \right]^{-1} \bgam_1~,
\eeq
and
\beq
\bA_n = - \left( \bbet_{n+1} + \balp_{n+1} \bA_{n+1} \right)^{-1} \bgam_{n+1}~,
\eeq
with ${}^{-1}$ denoting the matrix inverse. 
Non-trivial solutions $\bxi_0$ exist if 
\beq
\text{det} \left| \bM \right| = 0~. \label{eq:det}
\eeq

We used a numerical root finder to find pairs $\{\lambda, \omega\}$ that simultaneously satisfy Eq.~(\ref{eq:ctdfrac}) and Eq.~(\ref{eq:det}). As initial estimates for the root-finding algorithm, $\{\lambda_0, \omega_0\}$, we made use of the series expansions in Ref.~\cite{Dolan:2009kj} and the hydrogenic approximation, Eq.~(\ref{eq:omR-hydrogen}). 

\subsection{Time domain evolution\label{subsec:td-method}}

Several groups have investigated the excitation of bound states by generic initial data, using time-domain codes. The majority of work has focussed on bosonic fields, in the scalar \cite{Yoshino:2012kn, Dolan:2012yt, Yoshino:2013ofa, Yoshino:2014wwa, Burt:2011pv, Barranco:2012qs, Barranco:2013rua, Burt:2011pv} and Proca cases \cite{Witek:2012tr}. The evolution of Dirac bound states was investigated by Zhou \emph{et al.}~in Ref.~\cite{Zhou:2013dra}. 

We developed a 1+1D time-domain code to solve the coupled radial equations for the Schwarzschild case, written in the form,
\begin{align}
\frac{\partial F}{\partial t} &= \frac{\partial G}{\partial r_\ast} + \frac{\lambda f^{1/2}}{r} G + i \mu f^{1/2} F, \\
\frac{\partial G}{\partial t} &= \frac{\partial F}{\partial r_\ast} - \frac{\lambda f^{1/2}}{r} F - i \mu f^{1/2} G,
\end{align}
where $F = R_1 + R_2$ and $G = i(R_2 - R_1)$ and $\lambda = \ldots, -2, -1, 1, 2, \ldots$. 
We used the method of lines with a fourth-order Runge-Kutta integrator, and fourth-order finite differencing on spatial slices. To implement the absorbing boundary condition at the event horizon, we used the `Perfectly Matched Layer' method, which was previously applied in the black hole context in Ref.~\cite{Dolan:2012yt}. This entailed adding extra terms $- \gamma(r_\ast)F$ and $-\gamma(r_\ast) G$ to the equations above, where $\gamma(r_\ast)$ is a non-negative smooth function which is zero for $r_\ast > -150M$. At the far-field boundary we imposed a reflecting boundary condition by setting $F(r_{\text{max}}) = 0$. We moved this boundary to a large radius, $r_{\text{max}} \sim 4000M$, to suppress its effect on the excitation and evolution of low-lying bound states. To analyse the spectrum, we computed the `spectral power' 
\beq
P(\omega) =  |\tilde{F}|^2 + |\tilde{G}|^2 \label{eq:power}
\eeq 
from the square magnitudes of the (discrete) Fourier components $\tilde{F}(\omega)$ and $\tilde{G}(\omega)$.

\section{Results\label{sec:results}}

\subsection{Perturbation theory: $M \mu \ll 1$\label{subsec:pert-theory}}
Many authors have shown that, in the limit $M \mu \ll 1$, the bound state spectrum is hydrogenic; see Eq.~(\ref{eq:omR-hydrogen}). In this regime, the spectrum is degenerate, as it depends only on the principal quantum number $n = \ell + \hat{n} + 1$ (where $\ell$ is the orbital angular momentum [$j = \ell + S$, where $S \in \{-1/2, +1/2\}$] and $\hat{n} = 0, 1, \ldots$ is the excitation number), rather than on $j$, $\ell$ and $\hat{n}$ individually. 

How are degeneracies broken at higher orders in $M \mu$? The spectrum may be written as a series in $M \mu$, as follows,
\beq
\omega_R / \mu \approx 1 + \mathcal{E}_{n}^{(0)} + \mathcal{E}_{n \ell j}^{(1)} + \mathcal{E}_{n \ell j}^{(2)} + \ldots. 
\eeq
The `hydrogenic' term $\mathcal{E}_{n}^{(0)} = - \frac{(M\mu)^2}{2 n^2}$ is familiar; the `fine-structure' term $\mathcal{E}_{n \ell j}^{(1)} \sim O( (M\mu)^4 )$ and `hyperfine-structure' term $ \mathcal{E}_{n \ell j}^{(0)} \sim O\left( (M\mu)^5 (\tfrac{am}{M})\right)$ are less well-known \cite{Ternov:1980st, Ternov:1988xy, Dolan:2007:thesis}. Here we review these terms, and test against numerical data.

\subsubsection{Fine structure}
In the case of the hydrogen atom, the fine-structure correction arises at $O(\alpha^4)$ where $\alpha$ is the fine-structure constant, and it is given by \cite{Griffiths:1995} 
\beq
\mathcal{E}_{nl}^{(1,\text{Hyd})} = \frac{\alp^4}{n^4} \left(\frac{3}{8} - \frac{n}{2j+1} \right). \label{eq:hydrogen-fs}
\eeq
The second term, due to a spin-orbit coupling, breaks the degeneracy between states of the same orbital angular momentum but opposite spin projection (i.e.~the $j = \ell + 1/2$ and $j = \ell - 1/2$ states, for $\ell \ge 1$). 

The fine structure of the Schwarzschild bound state spectrum was calculated in Chap.~5 of Ref.~\cite{Dolan:2007:thesis}. The calculation is outlined in Appendix \ref{appendix:fine-structure}, where a small but vital correction is identified.  The key result is
\beq
\mathcal{E}_{j \ell n}^{(1)} = \frac{(M\mu)^4}{n^4} \left( \frac{15}{8} - \frac{3n}{2j+1} - \frac{3n}{2\ell + 1} \right) 
\label{eq:fine-structure} .
\eeq
The final term, not present in the hydrogen case (\ref{eq:hydrogen-fs}), implies that the bound-state spectrum depends on three quantum numbers individually: $j, \ell, n$ (rather than just $j$, $n$ as in Eq.~(\ref{eq:hydrogen-fs})).

\begin{figure}
\begin{center}
\includegraphics[width=12cm]{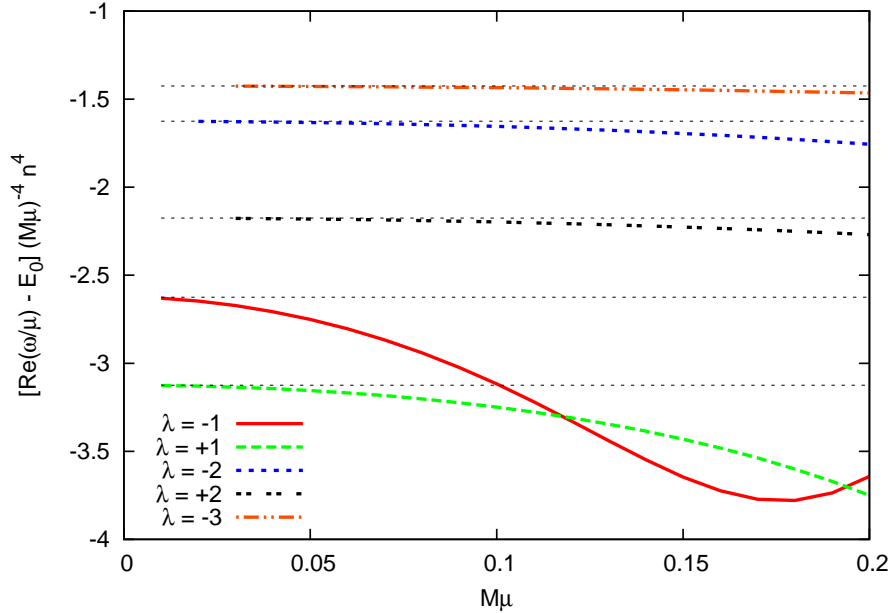}
\end{center}
\caption{
Fine structure of Dirac energy levels for Schwarzschild black hole. 
 The solid lines show numerical data for the difference $\text{Re}(\omega / \mu) - (1 - \frac{1}{2} (M\mu)^2 / n^2)$, where $n = \ell + \hat{n} + 1$ is the principal quantum number, with the y-axis scaled by $(M \mu)^{-4} n^4$. 
The thick lines show the ground-state modes ($\hat{n}=0$) with $\lambda = -1$ ($j=1/2$, $\ell = 0$), $\lambda = +1$ ($j =1/2$, $\ell = 1$), $\lambda = -2$ ($j = 3/2$, $\ell = 1$), $\lambda = +2$ ($j =3/2$, $\ell = 2$), $\lambda = -3$ ($j = 5/2$, $\ell = 2$). The horizontal dotted lines show the fine-structure prediction of Eq.~(\ref{eq:fine-structure}), with coefficients $-21/8$, $-25/8$, $-13/8$, $-87/40$ and $-57/40$, respectively.
}
\label{fig:fine-structure}%
\end{figure}

Figure \ref{fig:fine-structure} confronts Eq.~(\ref{eq:fine-structure}) with numerical data. Here, we plot the difference between $\omega_R / \mu$ and the hydrogenic result $1 + \mathcal{E}_{n}^{(0)}$, and we scale this difference by $n^4 (M\mu)^{-4}$. The plot shows excellent agreement between the prediction of Eq.~(\ref{eq:fine-structure}) (indicated by horizontal dotted lines) and the numerical data (thicker lines), for various modes. This agreement in the $M\mu \rightarrow 0$ regime increases our confidence in the validity of both the numerical method, and the perturbation-theory analysis that leads to Eq.~(\ref{eq:fine-structure}).

Many years ago, Ternov and Gaina \cite{Ternov:1988xy} also calculated fine-structure splittings for bound states, by applying standard perturbation theory techniques. Unfortunately, their fine-structure result at $O((M\mu)^4)$ is not found to be in agreement with Eq.~(\ref{eq:fine-structure}), nor with our numerical data. We suggest that the calculation in Ref.~\cite{Ternov:1988xy} is invalid due to the coordinate system singularity in Schwarzschild/Boyer-Lindquist coordinates, which causes physical quantities to diverge as $r \rightarrow r_+$. We note that, by contrast, Eq.~(\ref{eq:fine-structure}) was derived with a horizon-penetrating coordinate system. 

\subsubsection{Hyperfine structure}
In the hydrogen atom, the interaction between the magnetic dipole moments of the nucleus and electron leads to a spin-spin splitting of the $\ell = 0$ state, with the anti-aligned spin configuration lying at a lower energy. This is an example of `hyperfine' splitting, as it arises at subdominant order $\alp^4 (m_e / m_{\text{nuc.}})$. The hyperfine transition between $\ell = 0$ levels generates a 21cm neutral hydrogen line, of importance in astrophysics.

In Ref.~\cite{Ternov:1988xy}, it was calculated that the rotation of the black hole leads to a `hyperfine' splitting at $O\left( (M\mu)^5 (\tfrac{am}{M})\right)$. It was asserted therein that only the modes $\ell \ge 1$ are split in this way.

\begin{figure}
\begin{center}
\includegraphics[width=12cm]{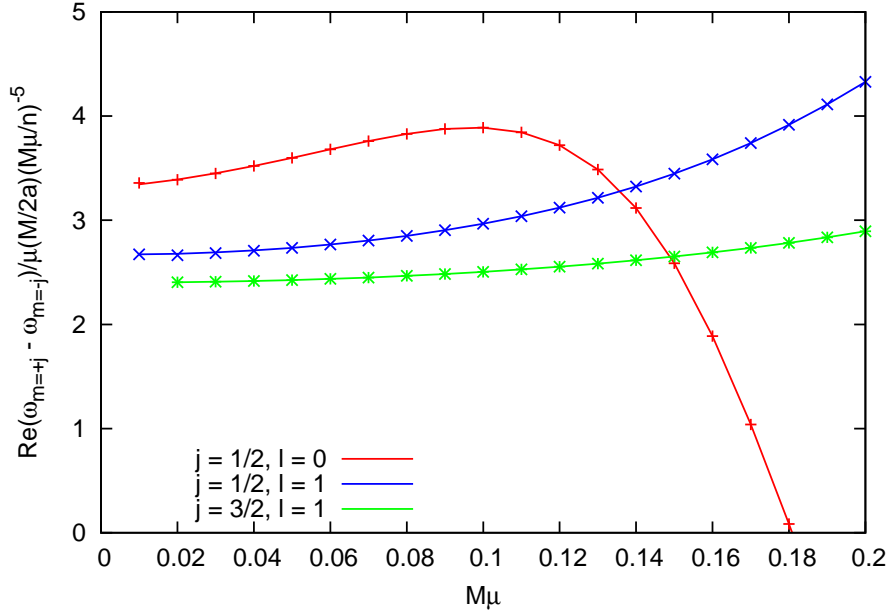}
\end{center}
\caption{
Hyperfine structure of Dirac energy levels for a slowly-rotating Kerr black hole. The plot shows numerical data for the difference in the energy of corotating and counter-rotating states ($m = +j$ and $m = -j$), after rescaling by $ \frac{1}{2} (M/a) n^5 (M\mu)^{-5}$. The solid lines show the $a=0.01M$ dataset, and the points show the $a=0.02M$ dataset; their agreement is evidence that the interaction is linear-in-$a$ at this order. The data supports the result of Ref.~\cite{Ternov:1980st} that rotation-induced hyperfine splitting scales with $(a m / M) (M \mu)^5$ at leading order (see text).
}
\label{fig:hyperfine}%
\end{figure}

In Fig.~\ref{fig:hyperfine} we attempt to extract the hyperfine correction from our numerical data, in the slow rotation regime ($a \le 0.02M$). The plot shows the difference between the energies of maximally co- and counter-rotating modes, after rescaling by $\frac{1}{2} (M/a) n^5 (M\mu)^{-5}$. We find that the co-rotating mode is more weakly bound than the counter-rotating mode, as expected. The numerical data suggests that the \emph{scaling} of Ternov \emph{et al.}~\cite{Ternov:1988xy} is indeed correct. However, we find that the $\ell = 0$ mode is also split at this order, and the numerical coefficients found in Eq.~(38) of Ref.~\cite{Ternov:1980st} are \emph{not} consistent with our data.

\subsubsection{Decay rates}
In Ref.~\cite{Ternov:1980st}, Ternov \emph{et al.}~used an asymptotic matching method to derive a key result for the imaginary part of the frequency of Dirac bound states on the Kerr-Newman spacetime in the regime $M \mu / j \ll 1$, $a m / M \ll 1$ (see Eq.~(29) in Ref.~\cite{Ternov:1980st}). In the Kerr case,
\beq
\omega_I / \mu \approx - \alpha_{j\ell n} \left( M \mu \right)^{4 + 2 \ell + 2S} ,
\label{eq:ternov-decay}
\eeq
where $S = \pm 1/2$ and
\begin{align}
\alpha_{j\ell n} &= \left( \frac{r_+ - r_-}{r_+ + r_-} \right)^{1 + 2\ell} 
\left( \frac{r_+ - r_-}{(2l+1)(r_+ + r_-)} \right)^{2S}
\frac{(n+\ell)!}{n^{4+2\ell} (2\ell)!(2\ell+1)!(n-\ell-1)!} \nn \\
 & \quad \quad \quad \times \prod_{p=1}^{j+1/2} \left[ 1 + \frac{4 \Gamma^2}{(p-1/2)^2} \right] ,
\end{align}
with $\Gamma = (2Mr_+ \mu - a m) / (r_+ - r_-)$.

\begin{figure}
\begin{center}
\includegraphics[width=12cm]{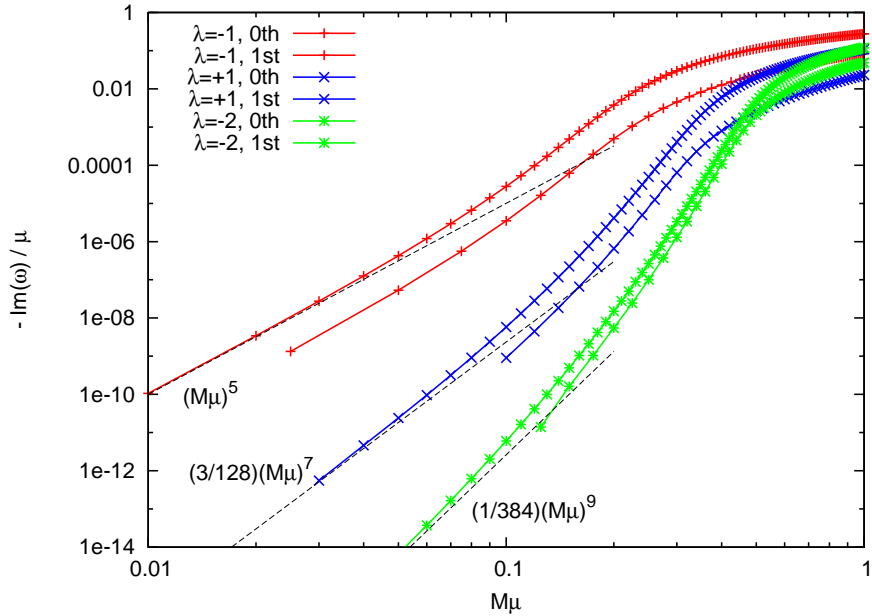}
\end{center}
\caption{
Power-law scaling of the decay rate of bound states on Schwarzschild spacetime. The solid lines show numerical data for $-\text{Im}(\omega / \mu)$ for the first two modes with $\lambda = -1$ ($j=1/2$, $\ell = 0$), $\lambda = +1$ ($j =1/2$, $\ell = 1$) and $\lambda = -2$ ($j = 3/2$, $\ell = 1$). The dotted lines show the asymptotic results $(M\mu)^5$, $\tfrac{3}{128} (M\mu)^7$ and $\tfrac{1}{384} (M\mu)^9$ (see Eq.~(\ref{eq:ternov-decay})). 
}
\label{fig:schw-decay}%
\end{figure}

Figure \ref{fig:schw-decay} shows numerical data for the imaginary part of bound states of the Schwarzschild black hole, for $n=1\ldots3$ modes. The plot illustrates how the decay rate has as a power-law scaling in the $M \mu \ll j$ regime. The leading-order results of Eq.~(\ref{eq:ternov-decay}) are shown as dotted lines. These results are found to be in good agreement with the data in the small-$M \mu$ regime. 

\subsection{Time-domain evolution\label{subsec:td-results}}
Are bound states typically excited by generic initial data? For the scalar field case, a number of studies \cite{Yoshino:2012kn, Yoshino:2013ofa, Yoshino:2014wwa, Dolan:2012yt, Burt:2011pv, Barranco:2012qs} have concluded that the answer is affirmative (see Ref.~\cite{Barranco:2013rua} for a Green's function analysis of the excitation factors in the initial-value formulation). Though the conclusion is expected to be similar in the Dirac case, it has received less attention \cite{Zhou:2013dra}, and we briefly investigate it here.

\begin{figure}
\begin{center}
\includegraphics[width=12cm]{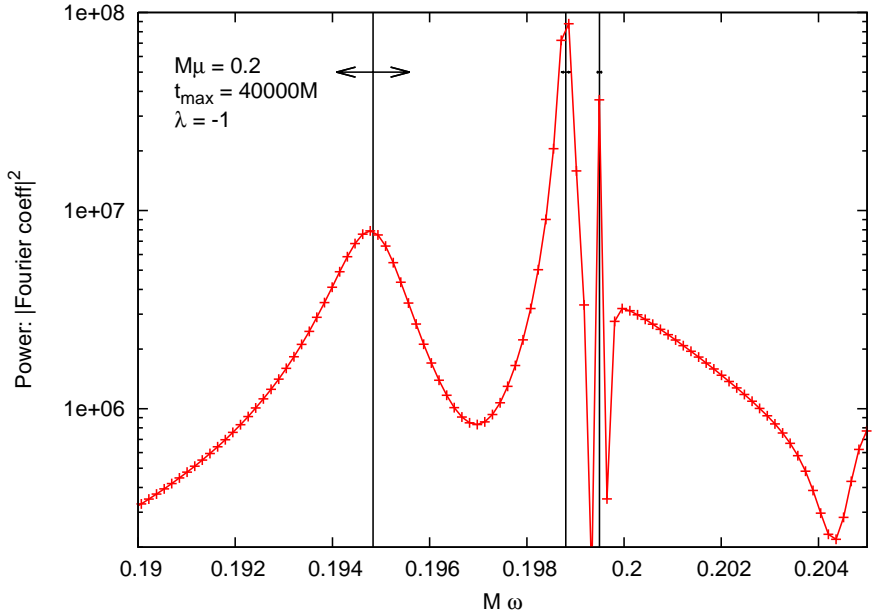}
\end{center}
\caption{
Power spectrum for a time-domain evolution of a massive Dirac field ($M\mu = 0.2$) on Schwarzschild spacetime ($a=0$). The $j=1/2, \ell = 0$ ($\lambda=-1$) mode was evolved with the 1+1D scheme outlined in Sec.~\ref{subsec:td-results}, starting with generic initial data described in Sec.~\ref{subsec:td-results}, up to $t_{\text{max}} = 4 \times 10^4 M$. The plot compares numerical data for the power [Eq.~(\ref{eq:power}), red symbols] with the spectral lines for the first three bound states [black vertical lines] determined via the frequency-domain method of Sec.~\ref{subsec:method-freq}). The expected Lorentzian half-widths are shown as horizontal arrows.
}
\label{fig:power}%
\end{figure}

Figure \ref{fig:power} shows the power spectrum $P(\omega)$ [Eq.~(\ref{eq:power})] resulting from a typical time-domain evolution, with parameters $M\mu = 0.2$, $a=0$, $\lambda=-1$, and Gaussian initial data $F(r) = 0$, $G(r) = \exp\left(-(r_\ast - r_\ast^{(0)})/(2 \sigma^2) \right)$, with $r_\ast^{(0)} = 90$ and $\sigma = 40$. There is clear evidence here that the first three bound states were excited by the initial data, at the expected frequencies ($\omega_R / \mu \approx 0.9741726$, $0.9940106$, and $0.9974677$). Furthermore, the `spectral lines' have the expected Lorentzian profiles, $P \sim 1/[(\omega-\omega_R)^2 + \omega_I^2]$. The widths are found to be in proportion to the imaginary parts of frequencies ($-\omega_I / \mu \approx 3.755 \times 10^{-3}$, $4.979 \times 10^{-4}$ and $1.407 \times 10^{-5}$). Consequently, the spectral line for the ground state is found to be the widest, as this mode decays most rapidly. 

\subsection{$M \mu \gg 1$ regime\label{subsec:semiclassical}}
In the `semi-classical' regime $M\mu \gg 1$, in which the gravitational length scale is much longer than the quantum-field length scale, we expect the key features of the spectrum to relate closely to the properties of timelike geodesics. Indeed, this quantum-to-classical correspondence emerges via a standard WKB analysis, as shown in Appendix \ref{sec:wkb}. 

For circular geodesics on the Schwarzschild spacetime, the (dimensionless) energy $\hat{\omega}$ is related to the angular momentum $\hat{L}$ and orbital radius $\hat{r}_0$ by:
\begin{align}
\hat{\omega} &= \sqrt{\left(1 - \frac{2}{\hat{r}_0} \right) \left(1 + \frac{\hat{L}^2}{\hat{r}_0^2} \right)}.   \label{eq:geodesicE}
\end{align}
The orbital radius $\hat{r}_0$ is given in terms of $\hat{L}$ in Eq.~(\ref{eq:r0}). For $\hat{L} \gg 1$, $\hat{\omega} \approx 1 - \tfrac12 \hat{L}^{-2} - \tfrac98 \hat{L}^{-4} + \ldots$. In Appendix \ref{sec:wkb}, it is shown that the geodesic parameter $\hat{L}$ should be associated with $(\ell + 1/2) / M\mu$ for a scalar field.

\begin{figure}
\includegraphics[]{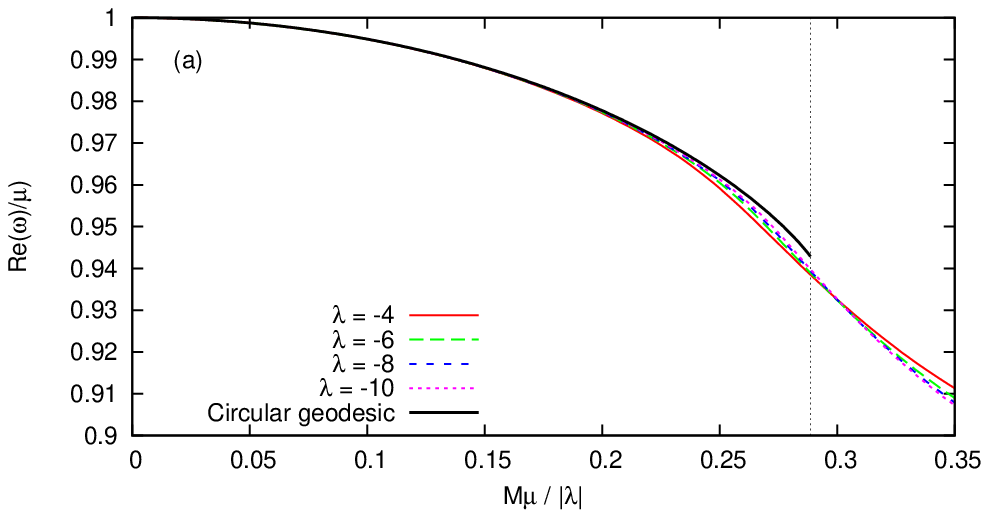}
\includegraphics[]{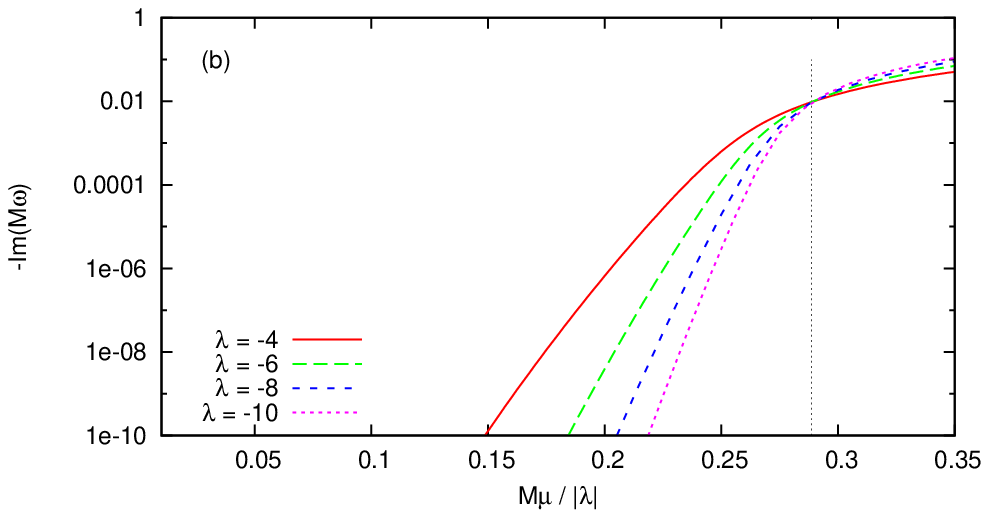}
\includegraphics[]{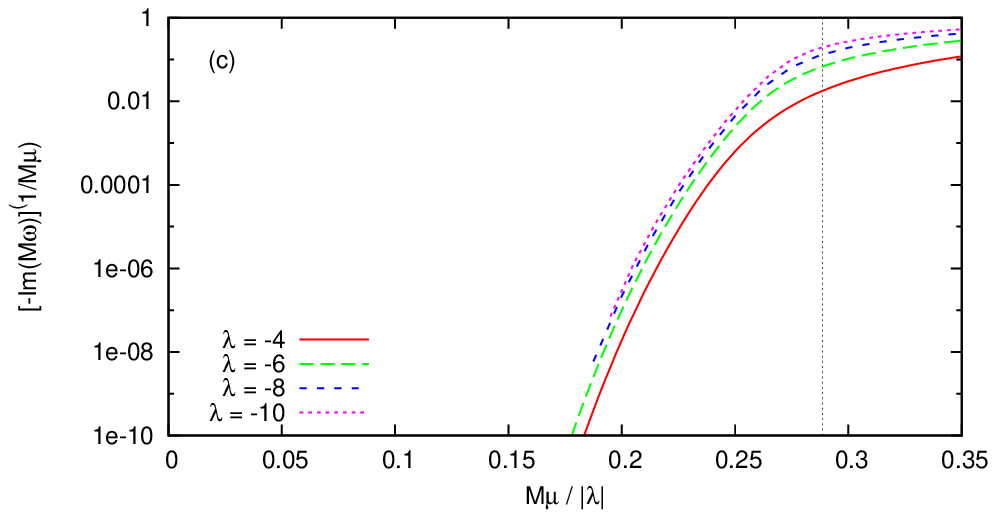}
\caption{Bound state frequencies in semi-classical regime ($M\mu \gg 1$, $|\lambda| \gg 1$), for a Schwarzschild black hole. The dotted line at $M\mu / |\lambda| = (12)^{-1/2}$ corresponds to the innermost stable circular orbit at $r=6M$.
}%
\label{fig:semiclassical}%
\end{figure}

Plot (a) of Fig.~\ref{fig:semiclassical} shows numerical data for the $\hat{n}=0$ energy level of a selection of angular modes in the range $-4 \ge \lambda \ge -10$. The energy level is shown as a function of $M \mu / |\lambda|$. We may associate this parameter with $1/\hat{L}$ in the classical limit. Making this association,  the geodesic energy level of Eq.~(\ref{eq:geodesicE}) is shown as a solid black line. The agreement between the data and geodesic prediction is evident. A dotted vertical line indicates the innermost stable circular orbit ($r_0 = 6M$, $\hat{L} = 1/\sqrt{12}$), beyond which stable circular orbits do not exist. Our numerical data suggests that, though bound-state solutions exist beyond this limit, they are rapidly-decaying.

Plots (b) and (c) of Fig.~\ref{fig:semiclassical} show the imaginary part of the frequency of these modes, as a function of $M \mu / |\lambda| \Leftrightarrow 1/\hat{L}$. Plot (b) shows that the decay rates are very similar at the position of the innermost stable circular orbit, $\hat{L} = \sqrt{12}$ (the point at which the angular momentum potential barrier separating the circular orbit from the horizon disappears). Plot (c) shows that the decay rate is consistent with 
\beq
\text{Im}(\omega / \mu) \sim \exp\left( -\beta( \hat{L} ) M \mu\right) 
\eeq
where $\beta( \hat{L} )$ is some quantum-tunnelling factor, determined from an integral across the potential barrier.

\subsection{Kerr bound states: decay\label{subsec:kerr-decay}}
\begin{figure}
\begin{center}
\includegraphics[width=8cm]{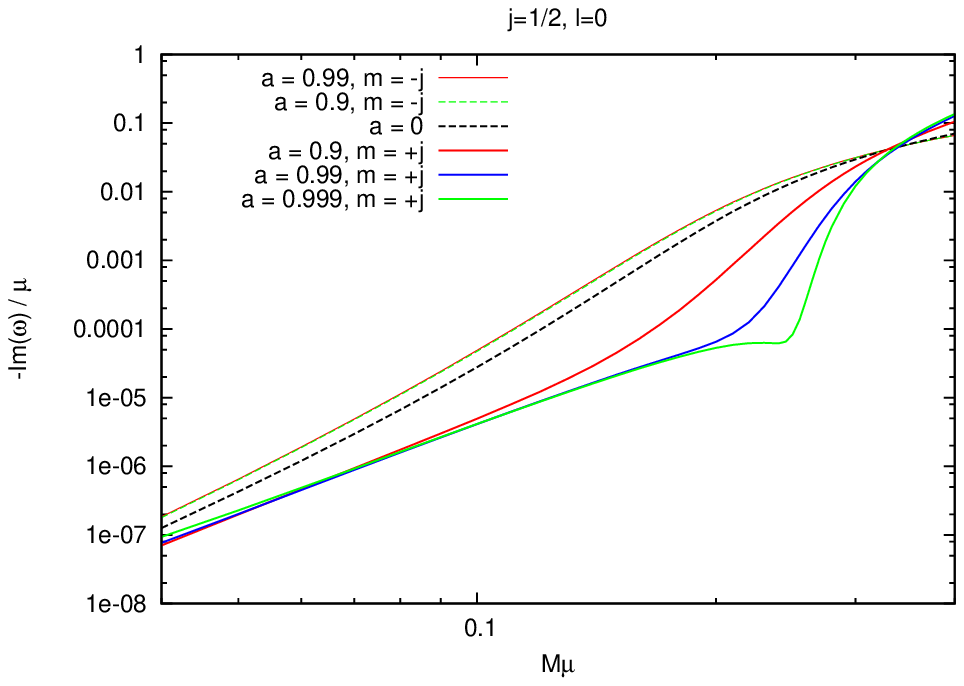}
\includegraphics[width=8cm]{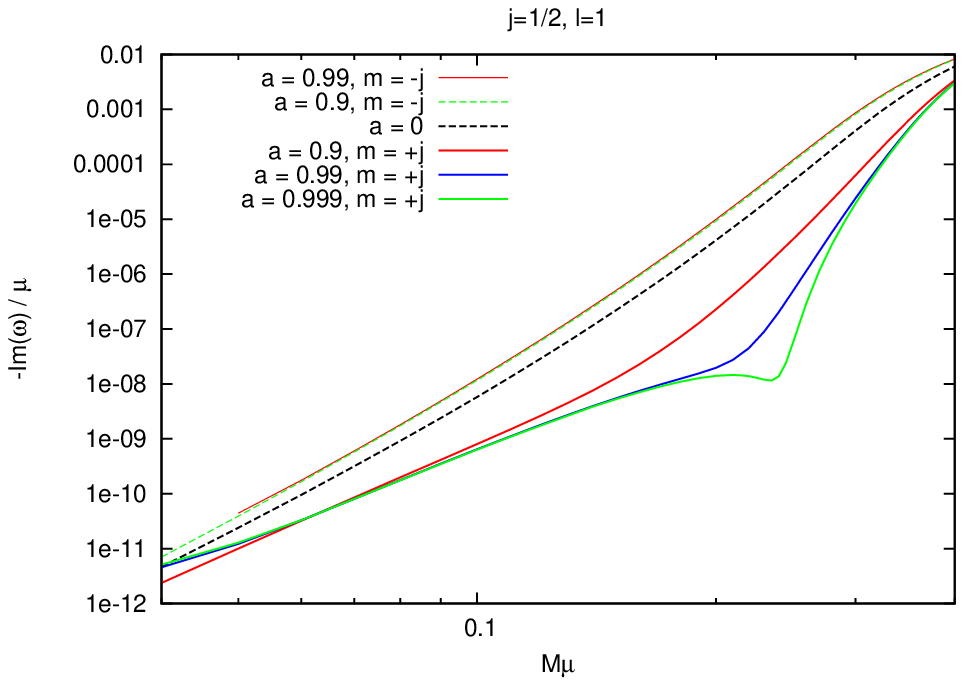}
\includegraphics[width=8cm]{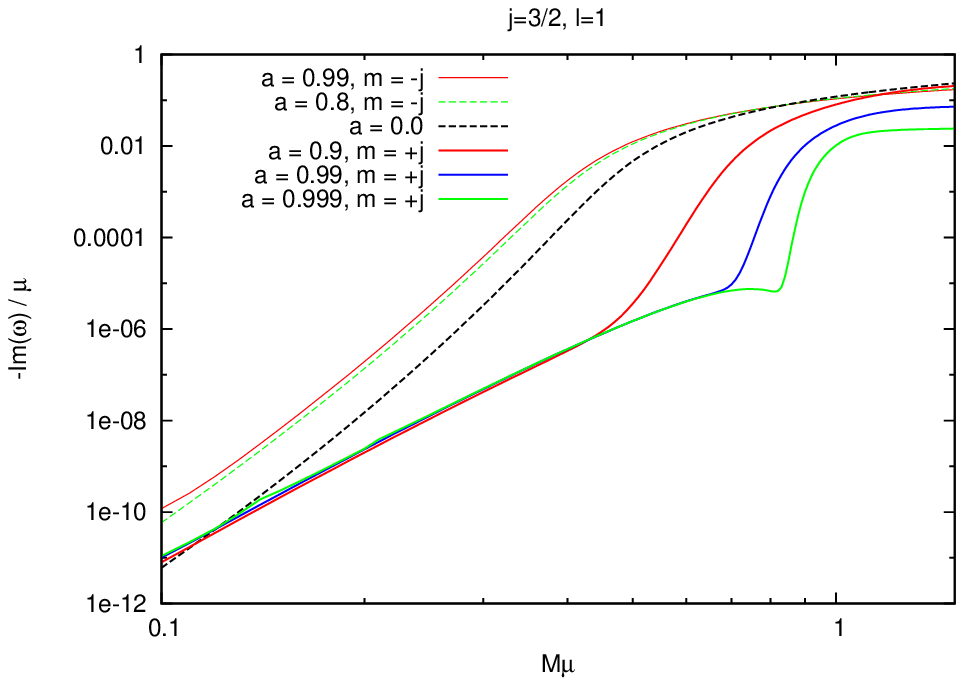}
\end{center}
\caption{
The decay rate of bound states on Kerr spacetime for co-rotating ($m=+j$) and counter-rotating ($m=-j$) modes.  The plots show $-\text{Im}(\omega / \mu)$ for the $j=1/2$, $\ell = 0$ [left] $j=1/2$, $\ell=1$ [right] and $j=3/2$, $\ell=1$ [lower] states, as a function of $M\mu$, on logarithmic axes. }
\label{fig:kerr-decay}%
\end{figure}

Now we focus attention on the decay rate of modes on a rotating black hole spacetime. Figure \ref{fig:kerr-decay} shows the imaginary part of bound state frequency as a function of $M\mu$, for both co-rotating ($m > 0$) and counter-rotating ($m < 0$) cases. For low couplings $M\mu \ll 1$, we see that the counter-rotating case is rather similar to the Schwarzschild case: the decay rate is governed by a power law, in accord with Eq.~(\ref{eq:ternov-decay}). 

The co-rotating modes exhibit an interesting feature. In the regime $\omega < m \Omega_H$ the decay rate is suppressed; whereas outside this regime, the decay rate increases rapidly. The plots suggest that, for rapidly-rotating black holes $a \gtrsim 0.99$, the decay rate has a local minimum precisely at $\omega = m \Omega_H$. 

\begin{figure}
\begin{center}
\includegraphics[]{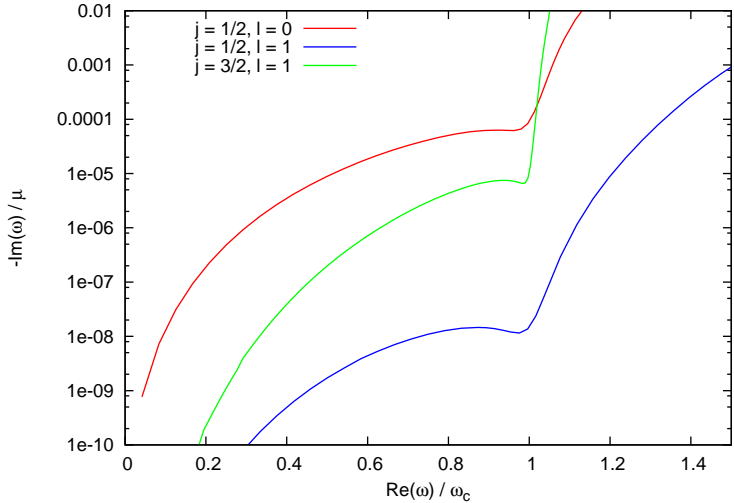}
\end{center}
\caption{
Decay rate for co-rotating modes $m = +j$ at $a = 0.999M$ as a function of $\omega / \omega_c$, where $\omega_c = m \Omega_H$ is the critical superradiant frequency. The plot shows that decay is suppressed for $\omega \tilde{\omega} < 0$, and that a local minimum arises at $\omega = \omega_c$. 
}
\label{fig:omegac}%
\end{figure}

Figure \ref{fig:omegac} shows the decay rate as a function of $\omega / m \Omega_H$, where $\Omega_H$ is the angular frequency of the event horizon. The plot makes it evident that the local minimum coincides with the `critical' superradiant frequency $\omega_c = m \Omega_H$. The data suggests that minimum value decreases with increasing $a$, which hints at the possibility that there may be a non-decaying `critical' mode in the extremal case $a \rightarrow M$. This possibility remains to be investigated.

\subsection{Kerr bound states: spectrum}

Now we turn attention to results for the Kerr energy spectrum. In the limit $M \mu \rightarrow 0$ the spectrum is well-understood, as outlined in Sec.~\ref{subsec:pert-theory}, but outside this regime the spectrum has not been examined in detail.

\begin{figure}
\begin{center}
\includegraphics[]{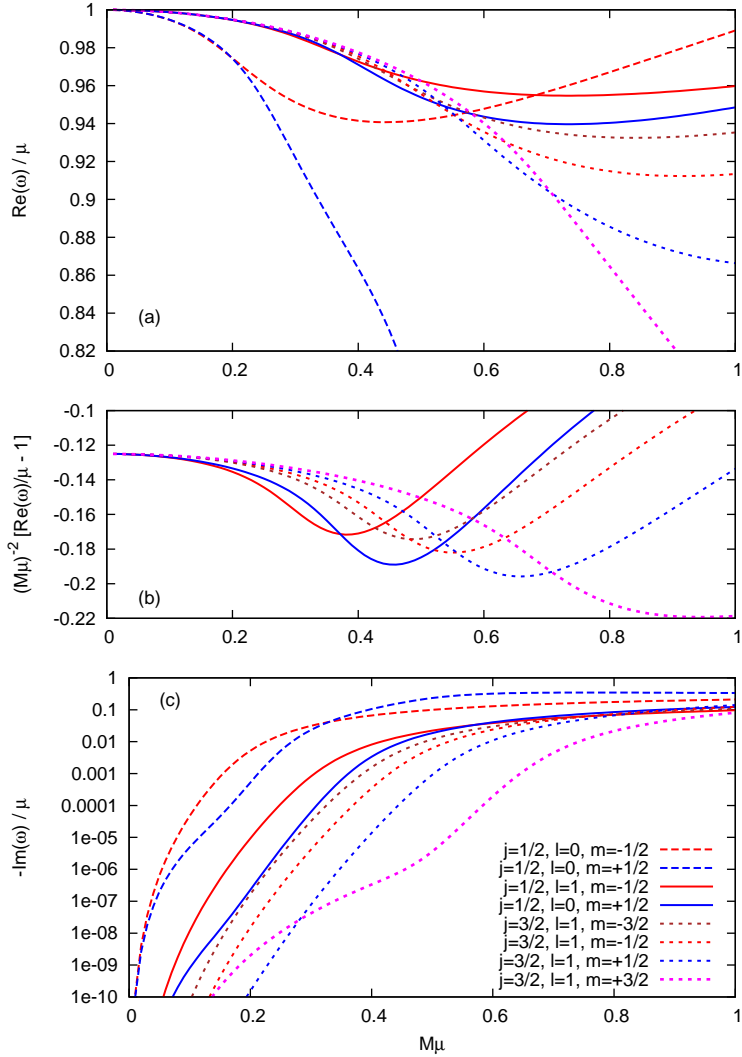}
\end{center}
\caption{
Spectrum of $n=1$ and $n=2$ Dirac bound states for Kerr black hole at $a = 0.9M$. Plot (a) shows the real part of the frequency, $\text{Re}(\omega) / \mu$, as a function of $M\mu$, for $j=1/2$, $\ell = 0$ (dashed), $j=1/2$, $\ell =1$ (solid) and $j=3/2$, $\ell=1$ (dotted) modes. Plot (b) shows the fine structure, $[\text{Re}(\omega) / \mu - 1] / (M\mu)^2$, for the $n=2$ states. At low $M \mu$, the $j=1/2$, $\ell = 1$ states [solid] are lower-lying than the $j = 3/2$, $\ell = 1$ states [dotted]; and the counter-rotating states ($m < 0$) are lower-lying than the co-rotating states ($m > 0$). The ordering changes as $M\mu$ increases, as the imaginary part of the frequency increases: see plot (c). 
}
\label{fig:spectrum-a9}%
\end{figure}

Figure \ref{fig:spectrum-a9} shows the spectrum of the lowest modes $n=1$ and $n=2$ (where $n$ is the principal quantum number) for a Kerr black hole with $a=0.9M$, as a function of $M\mu$. Let us make several observations. (1) At low $M\mu$, the $j=1/2$ co-rotating ($m=+1/2$) state is higher-lying, and decays more slowly, than the counter-rotating state ($m=-1/2$). (2) As $M\mu$ increases, the roles are reversed, with the co-rotating state lower-lying for $M\mu \gtrsim 0.2$ and faster-decaying for $M\mu \gtrsim 0.33$. (3) At higher couplings, the $j=1/2$ states appear to achieve large binding energies; but here the states are highly transient (with no classical analogue) and may not be physically significant. (4) At low $M\mu$, the $\ell = 1$ states are ordered according to the fine and hyperfine structure equations: the $j=\ell - 1/2$ states [solid] are lower-lying and faster-decaying than the $j = \ell + 1/2$ states [dotted], and the counter-rotating states ($m < 0$) are lower-lying than the co-rotating states ($m > 0$). Fig.~\ref{fig:spectrum-a9}(b) shows how this hierarchy changes as $M\mu$ increases, and the states become more transient. (5) The maximally-corotating modes (e.g.~$m=j=3/2$, $\ell=1$) are slowly-decaying in the regime $\text{Re}(\omega) < m \Omega_H$.

In Table \ref{tbl:spectrum_a9} we present a sample of numerical data for bound state frequencies, together with the corresponding angular separation constants.

\begin{table}
\begin{center}
\begin{tabular}{c c | c c | c c}
\hline\hline
Mode & $m$ & $\text{Re}(\omega)/\mu$ & $\text{Im}(\omega)/\mu$ & $\text{Re}(\lambda)$ & $\text{Im}(\lambda)$ \\
\hline
$j=1/2$, $\ell = 0,$	\quad & $-1/2$ 
&  $0.95120580$   &   $-3.1293497 \times 10^{-2}$  &  $-1.2612238$  &  $+5.6180382 \times 10^{-3}$ \\
 & $+1/2$ & 
$0.92200086$    &  $-2.3193867 \times 10^{-2}$  & $-0.7440728$  & $-4.1963431 \times 10^{-3}$ \\
\hline
$j=1/2$, $\ell = 1,$	\quad & $-1/2$ 
& $0.98580062$   &  $-8.1786335 \times 10^{-4}$  &  $+1.1074300$   & $-1.6229636 \times 10^{-4}$ \\
 & $+1/2$ 
 & $0.98662631$ &  $-6.3064813 \times 10^{-5}$ & $+0.9348739$  &$+9.8118563 \times 10^{-6}$ \\
\hline
$j=3/2$, $\ell = 1,$	\quad & $-3/2$ 
& $0.98744302$ & $-3.2293333 \times 10^{-5}$ & $-2.2672880$ & $+6.9736117 \times 10^{-6}$ \\
 & $-1/2$ & 
$0.98767652$  &  $-4.1491804 \times 10^{-6}$ & $-2.1091337$ & $+3.7495425 \times 10^{-7}$ \\
 & $+1/2$ &
 $0.98785317$ & $-7.7639988 \times 10^{-8}$ & $-1.9333717$ &   $-3.6997312 \times 10^{-9}$ \\
 & $+3/2$ & $0.98798850$ & $-4.2000702 \times 10^{-8}$ & $-1.7325948$ &  $-9.0746837 \times 10^{-9}$ \\
\hline \hline
\end{tabular}
\end{center}
\caption{
Sample numerical data for Dirac bound states ($n=1,2$) for a Kerr black hole at $M\mu = 0.3$, $a = 0.9M$. The columns show the real and imaginary parts of the frequency, $\omega / \mu$, and the angular separation constant, $\lambda$. 
}\label{tbl:spectrum_a9}
\end{table}


\subsection{Kerr bound states: wavefunctions\label{subsec:wf}}



The components of the Dirac current in the `ingoing Kerr' coordinate system were considered in Sec.~\ref{subsec:dirac-current}. In obtaining (\ref{eq:Jt-bl})--(\ref{eq:Jt-ingoing}), we tacitly assumed that the frequency $\omega$ is real. For bound states as defined in Sec.~\ref{sec:methods}, for which this is not the case, Eq.~(\ref{eq:Jt-ingoing}) should be multiplied by an additional factor of $\exp(2 \text{Im}(\omega) \tilde{t})$. After integrating over the two-surface of constant $r$, and applying the normalization condition (\ref{eq:Snorm}), we have
\beq
 \oint \rho^2 \tilde{J}^{\tilde{t}} d \Omega = e^{2 \text{Im} (\omega) \tilde{t}} \left( \Upsilon_{(0)} + \Upsilon_{(1)}  \right) 
\eeq 
where
\beq
\Upsilon_{(0)} = \left( \frac{(r - r_+)^{r_+}}{(r - r_-)^{r_-}} \right)^{-2 \text{Im}(\omega)} \,\left(|R_2|^2 + \left(r^2 + 2Mr + a^2 \right) |\Delta^{-1/2}R_1|^2 \right) \label{eq:upsilon}
\eeq
with $\Upsilon_{(1)}$ associated with the second term in Eq.~(\ref{eq:Jt-ingoing}). These terms are finite at the outer horizon.  

In the limit $M\mu \rightarrow 0$, the radial solutions are hydrogenic in form \cite{Lasenby:2002mc} (see Appendix \ref{appendix:fine-structure}). The radial functions may be written in terms of Laguerre polynomials in the dimensionless variable $x = (M\mu)^2 (r / M) n^{-1}$. The radial function has $\hat{n}$ nodes, where $\hat{n} = n - \ell - 1$ is the excitation number. 
  
\begin{figure}
\begin{center}
\includegraphics[]{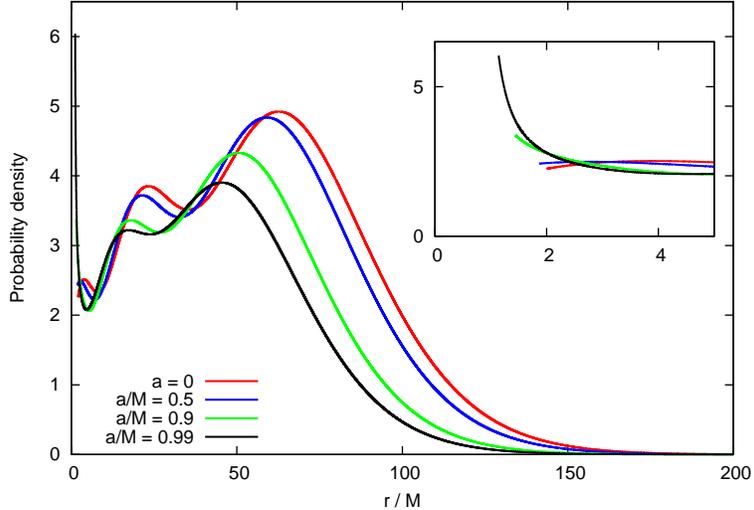}
\end{center}
\caption{
Profile of probability density from Dirac current, for $j=1/2$, $n=3$ bound states on Kerr black holes with $M\mu = 0.38$ and $a/M= 0$, $0.5$, $0.9$ and $0.99$. The plot shows $\Upsilon_{(0)}$, defined in Eq.~(\ref{eq:upsilon}), which is related to the Dirac probability density in the ingoing Kerr coordinate system. The inset shows that the density is finite at the outer event horizon, as expected.
}
\label{fig:density}%
\end{figure}
 
Figure~\ref{fig:density} shows the radial profile of $\Upsilon_{(0)}$ for the $j=m=1/2$, $\ell=0$ mode at $M\mu = 0.38$, for a variety of Kerr parameters $a$. Changing $a$ modifies the shape of the profile somewhat. In this case, the peaks in the profile move closer to the horizon, and the value on the horizon also increases, as shown in the inset.

\section{Discussion and Conclusions\label{sec:conclusions}}
In this article we have explored the spectrum of bound states of the massive Dirac equation on Kerr spacetime: 
 `trapped' modes which are ingoing at the horizon and fall off exponentially towards spatial infinity.
Let us now briefly review the key results. 

In Sec.~\ref{sec:formalism} we formulated the Dirac equation on Kerr spacetime, reviewing (i) the separation of variables, (ii) the absence of superradiance, and (iii) the violation of the weak energy principle. Eschewing the Newman-Penrose 2-spinor formalism used in pioneering works \cite{Chandrasekhar:1976ap, Chandrasekhar:1983, Page:1976jj}, we favoured instead the 4-spinor formalism \cite{Brill:1957fx, Unruh:1973, Unruh:1974bw, McKellar:1993ej, Finster:1999ry, Casals:2012es}, with the Weyl representation; positive spacetime signature; and Carter's canonical (`symmetric') tetrad. 

In Sec.~\ref{sec:methods} we presented a practical method for computing the bound state spectrum, reducing the problem to that of finding the roots of certain (matrix-valued) three-term recurrence relations. We also briefly described a time-domain method for the Schwarzschild case, with scope for extension to the Kerr case. 

In Sec.~\ref{subsec:pert-theory} we confronted small-$M\mu$ asymptotic results, obtained in the 1980s by Ternov, Gaina and coworkers \cite{Ternov:1988xy}, with new numerical data. We concluded that (i) the decay of bound states is well described by Eq.~(\ref{eq:ternov-decay}); and (ii) the \emph{scaling} of the fine-structure and hyperfine-structure terms has been correctly deduced. However, the numerical coefficients for the (hyper)fine-structure terms were found to be inconsistent with the data. We found that, instead, our data is fully consistent with the fine-structure result presented in Eq.~(\ref{eq:fine-structure}), which was derived in Ref.~\cite{Dolan:2007:thesis} and Appendix \ref{appendix:fine-structure}.

In Sec.~\ref{subsec:td-results} we used a time-domain code to demonstrate that bound states are typically excited by generic initial data. In Sec.~\ref{subsec:semiclassical} we briefly considered the large-$M\mu$ regime, showing in Fig.~\ref{fig:semiclassical} that our new results are consistent with `semi-classical' expectations from a WKB analysis (e.g.~Appendix \ref{sec:wkb}). In Secs.~\ref{subsec:kerr-decay}--\ref{subsec:wf} we examined the novel features of bound states of rapidly-rotating black holes. We found that the decay rate of the maximally-corotating mode is strongly suppressed in the regime $\omega < m \Omega_H$ (Figs.~\ref{fig:kerr-decay}--\ref{fig:omegac}). 

Let us now discuss some of the implications of our findings. For the \emph{scalar} field, modes precisely at the superradiant transition frequency ($\omega_R = m \Omega_H$) are stationary (i.e.~non-decaying) \cite{Detweiler:1980uk, Hod:2012px, Hod:2013zza}. Herdeiro and Radu have recently shown that this transition mode  is implicated in the Kerr family `branching off' into a new family of `hairy' solutions with (complex massive) scalar-field hair \cite{Herdeiro:2014goa, Benone:2014ssa, Herdeiro:2015gia}. In the Dirac case, the transition mode is \emph{not} stationary, due to the lack of superradiance. On the other hand, as decay may be strongly suppressed (Fig.~\ref{fig:omegac}), it is plausible that rapidly-rotating black holes can support very long-lived co-rotating Dirac hair.

Let us now highlight several avenues for future work which could lead to a more complete understanding. Despite the progress described above, this work lacks: (i) correct coefficients for the hyperfine structure at $O( (a m/M) (M \mu)^5)$; (ii) an asymptotic result for the rate of decay for co-rotating modes for $\omega \le m \Omega_H$, $a \lesssim M$; (iii) asymptotic WKB results at \emph{subleading} order (for $M\mu \gg 1$) to account for the effects of spin. Revisiting the asymptotic analysis of the Dirac equation in second-order form \cite{Ternov:1980st, Gal'tsov:1983, Ternov:1986bf, Gaina:1988dp, Ternov:1988xy} could pay dividends.

Another possibility would be to examine the bound-state wavefunction \emph{inside} the black hole horizon. It is well-known that the inner (Cauchy) horizon at $r=r_-$ has the effect of repelling geodesics. Therefore, it seems unlikely that the bound states investigated here will be `ingoing' at $r=r_-$, in general.  Nevertheless, there may exist particular frequencies for which the mode may be ingoing at both horizons. This remains to be investigated.

A further topic for investigation is the excitation of Dirac bound states by generic initial data on Kerr spacetime. Two approaches suggest themselves. First, an extension of the Green's function analysis of Ref.~\cite{Barranco:2013rua} to the Dirac case. Second, an extension of the time-domain approach of Sec.~\ref{subsec:td-method} and Refs.~\cite{Dolan:2012yt, Zhou:2013dra} to the $a > 0$ case.

In this work we have treated the Dirac field as a `classical' field, which is of course not the case. Unruh's  second-quantized analysis \cite{Unruh:1974bw} (1974) showed that fermions, as well as bosons, experience an instability to spontaneous particle creation (Unruh-Starobinski radiation \cite{Starobinskii:1973, Unruh:1974bw}); that is, a quantum version of superradiance. In 1983, Gal'tsov \emph{et al.}~\cite{Gal'tsov:1983} considered the filling of Dirac bound states in the Schwarzschild case, concluding that the distribution is thermal ($N \approx [8(1 + \exp(\omega / \kappa T_H))]^{-1}$) in the limit $M\mu \ll 1$, where $T_H$ is the Hawking temperature. The filling of states on the Schwarzschild spacetime was also considered in Ref.~\cite{Grain:2007gn}. Hartman \emph{et al.}~\cite{Hartman:2009qu} have considered the Kerr case, arguing that \emph{all} bound states with energies $\omega < m \Omega_H$ will be filled in Unruh's vacuum, creating a (stable) `Kerr-Fermi sea' which extends outside the ergosphere. This intriguing possibility undoubtedly deserves some further consideration.

\acknowledgments
With thanks to Helvi Witek and Paolo Pani. The work of S.D.~was supported in part by the Lancaster-Manchester-Sheffield Consortium for Fundamental Physics under STFC grant ST/L000520/1, and by EPSRC grant EP/M025802/1.

\appendix

\section{WKB analysis of Schwarzschild bound states\label{sec:wkb}}
Consider a scalar field $\Phi$ on Schwarzschild spacetime, satisfying $\Box \Phi - \mu^2 \Phi = 0$, which may be decomposed into modes in the standard way,
\beq
\Phi = \frac{1}{r} e^{-i \omega t} u(r) Y_{lm}(\theta, \phi) .
\eeq
The radial equation is
\beq
\left\{ \frac{d^2}{dr_\ast^2} + \omega^2 - V_0(r) \right\} u = 0, \quad \quad V_0(r) = f(r) \left(\mu^2 + \frac{l(l+1)}{r^2} + \frac{2M}{r^3} \right), \label{eq:scalar}
\eeq
where $f(r) = 1 - 2M/r$ and $r_\ast = r + 2M\ln(r/2M -1)$. 
In the regime $M\mu \gg 1$ (i.e.~$r_h \gg \lambda_c$), this may be written in dimensionless form as  
\beq
\left\{ \frac{d^2}{d\hat{r}_\ast^2} + (M \mu)^2 \, \hat{U}(\hat{r}) \right\} u = 0, \quad \quad 
\hat{U}(\hat{r}) = \hat{\omega}^2 - \hat{V}(\hat{r})  + O\left((M\mu)^{-2}\right)
, \label{eq:schro}
\eeq
and
\beq
\hat{V}(\hat{r}) = f(r) \left(1 + \frac{\hat{L}^2}{\hat{r}^2}\right) 
\eeq
where $\hat{r} = r/M$, $\hat{r}_\ast = r_\ast / M$, $\hat{\omega} = \omega / \mu$ and $\hat{L} = (\ell + 1/2) / (M\mu)$. Note that (\ref{eq:schro}) resembles a Schr\"odinger equation with a large parameter $M \mu \Leftrightarrow \sqrt{2m} / \hbar$, and thus we may apply standard WKB methods to reach the Bohr-Sommerfeld condition for bound states:
\beq
\int \hat{U}^{1/2} d\hat{r}_\ast = \frac{2 \pi}{M\mu} (n + 1/2) , \quad \quad n \in \mathbb{N}. \label{eq:bohr-sommerfeld}
\eeq

At this point we note the close analogue between $U(\hat{r})$ in (\ref{eq:schro}) and the right-hand side of the geodesic `energy' equation,
\beq
\dot{r}^2 = U_{\text{geo}}(r) \equiv \mathcal{E}^2 - f(r) \left(1 + \frac{\mathcal{L}^2}{\hat{r}^2}\right) 
\eeq
where $\mathcal{E} = f \dot{t}$ and $\mathcal{L} = r^2 \dot{\phi} / M$ are constants of motion, and the overdot notation indicates differentiation with respect to proper time. Circular orbits are defined by $dU_{\text{geo}}/d\hat{r} = 0$ with stable orbits satisfying $d^2U_{\text{geo}}/dr^2 < 0$. The former condition leads to an equation for the circular orbit radius in terms of the dimensionless angular momentum: 
\beq
\hat{r}_0 = \tfrac{1}{2} \hat{L}^2 \left( 1 + \sqrt{1 - 12/\hat{L}^2} \right) .  \label{eq:r0}
\eeq 
The latter condition implies that $\hat{r}_0 \ge 6$ and $\hat{L}^2 > 12$ for stable orbits. After expanding around the circular-orbit radius, and inserting into (\ref{eq:bohr-sommerfeld}), we obtain
\beq
\hat{\omega}^2 = V_0 +  \sqrt{2 V''} f(r_0) \frac{(n+1/2)}{M\mu} + \ldots ,
\eeq
where $V_0 = V(\hat{r}_0)$ and $V'' = \left. \frac{d^2 V}{d \hat{r}^2} \right|_{\hat{r}_0}$. Equivalently,
\beq
\hat{\omega} = V_0^{1/2} + \sqrt{\frac{V''}{2 V_0}} f(r_0)  \frac{(n+1/2)}{M\mu}  + \ldots
\eeq
In the weak-field regime, where $\hat{r}_0 \approx \hat{L}^2 \gg 1$, $V_0 \approx 1 - \hat{r}_0^{-1}$, $V'' \approx 2 / \hat{r}^3$ we reach Eq.~(\ref{eq:intro-wkb}) (after reinserting dimensionful constants).




\section{Carter tetrad and spin connection\label{sec:spinconnection}}
The inverse components of Carter's tetrad are
\begin{align}
e^0_\mu dx^\mu =& \frac{\sqrt{\Delta}}{\rho} \left(dt - a \sin^2 \theta d\phi \right) , & e^1_\mu dx^\mu =& \frac{\rho}{\sqrt{\Delta}} dr , \nn \\
e^3_\mu dx^\mu =& \frac{\sin \theta}{\rho} \left(- a dt + (r^2+a^2) d \phi \right) , & e^2_\mu dx^\mu =& \rho d \theta ,  \label{eq:canonical}
\end{align}

Noting the anti-symmetry relation $\omega_{\mu a b} = -\omega_{\mu b a}$, the non-trivial components of the spin connection are  
\begin{align}
\omega_{t\,01} &= -\frac{M(r^2 - a^2 \cos^2 \theta)}{\rho^4} , &
\omega_{t\,23} &= \frac{2 a M r \cos \theta}{\rho^4} , \nn \\
\omega_{r\,03} &= \frac{a r \sin \theta}{\rho^2 \sqrt{\Delta}} , &
\omega_{r\,12} &= -\frac{a^2 \sin \theta \cos \theta}{\rho^2 \sqrt{\Delta}} , \nn \\
\omega_{\theta\,03} &= -\frac{a \sqrt{\Delta} \cos \theta}{\rho^2} , &
\omega_{\theta\,12} &= -\frac{r \sqrt{\Delta}}{\rho^2} , \nn \\
\omega_{\phi\,01} &= \frac{a \sin^2\theta}{\rho^4} \mathcal{B} , &
\omega_{\phi\,02} &= \frac{a \sqrt{\Delta} \sin \theta \cos \theta}{\rho^2} , \nn \\
\omega_{\phi\,13} &= -\frac{r \sqrt{\Delta} \sin \theta}{\rho^2}, &
\omega_{\phi\,23} &= -\frac{\cos\theta}{\rho^4} \mathcal{A} ,
\end{align}
where $\mathcal{A} = \rho^2 \Delta + 2 M r (r^2 + a^2)$ and $\mathcal{B} = a^2 r \cos^2 \theta - a^2 M \cos^2 \theta + r^3 + M r^2$.  

The matrices $\Gamma_\mu$, defined in Eq.~(\ref{eq:Gam}), are given by
\begin{align}
\Gamma_t &= \frac{M}{2} 
\begin{pmatrix} 
\varrho^{-2} \, \sig_3 & O \\ 
O &  -{\varrho^\ast}^{-2} \, \sig_3
\end{pmatrix} , \nn \\
\Gamma_r &= -\frac{1}{2} \left(\frac{a \sin \theta}{\sqrt{\Delta}}\right)
\begin{pmatrix}
\varrho^{-1} \, \sig_2 & O \\ O & -{\varrho^\ast}^{-1} \, \sig_2
\end{pmatrix} , \nn \\
\Gamma_\theta &= -\frac{1}{2} \left(\sqrt{\Delta}\right) 
\begin{pmatrix}
(i\varrho)^{-1} \, \sig_2 & O \\ O & -{(i\varrho)^\ast}^{-1} \, \sig_2
\end{pmatrix} , \nn \\
\Gamma_\phi &= \frac{1}{2} \sqrt{\Delta} \sin \theta 
\begin{pmatrix}
(i\varrho)^{-1} \, \sig_1 & O \\ O & -{(i\varrho)^\ast}^{-1} \, \sig_1
\end{pmatrix} 
+ \frac{1}{2} 
\begin{pmatrix}
 \varpi \, \sig_3 & O \\ O & -\varpi^\ast \, \sig_3
\end{pmatrix} ,
\label{eq:Gamma_mu}
\end{align}
where $\varpi = i \cos \theta - a \varrho^{-2} (\varrho + M) \sin^2 \theta$.

\section{Connection with Finster \emph{et al.}\label{sec:finster}}
After multiplying Eq.~(\ref{eq:etapm}) by $\sigma_3$, we may write once more in the four-matrix form
\beq
\left( \mathcal{R}(r) + \mathcal{Q}(\theta) \right) \begin{pmatrix} \eta^- \\ \eta^+ \end{pmatrix} = 0
\eeq
where
\beq
\mathcal{R}(r) = 
\begin{pmatrix} 
 i \mu r & 0 & \sqrt{\Delta} \mathcal{D}_+ & 0 \\
 0 & -i \mu r & 0 & \sqrt{\Delta} \mathcal{D}_- \\
 \sqrt{\Delta} \mathcal{D}_- & 0 & -i \mu r & 0 \\
 0 & \sqrt{\Delta} \mathcal{D}_+ & 0 & i \mu r
 \end{pmatrix} ,
\quad
\mathcal{Q}(r) = 
\begin{pmatrix}
a \mu \cos \theta & 0 & 0 & \mathcal{L}_+ \\
0 & -a \mu \cos \theta & -\mathcal{L}_- & 0 \\
0 & \mathcal{L}_+ & a \mu \cos \theta & 0 \\
-\mathcal{L}_- & 0 & 0 & -a\mu \cos \theta 
\end{pmatrix} .
\eeq
with 
\begin{align}
\mathcal{D}_\pm &= \partial_r \pm \frac{i}{\Delta} \left[ -\omega (r^2+a^2) + a m \right] \\
\mathcal{L}_\pm &= \partial_\theta + \tfrac12 \cot \theta \pm \left(m \csc \theta - a \omega \sin \theta\right) .
\end{align}
A spinor transformation takes this to the form found in Refs.~\cite{Finster:1999ry, Finster:2000jz, Finster:2001vn, Finster:2008bg}.

\section{Fine-structure calculation for Schwarzschild bound states\label{appendix:fine-structure}}

Here we give an overview of a calculation in Chap.~5 in Ref.~\cite{Dolan:2007:thesis}, which leads to the fine-structure result, Eq.~(\ref{eq:fine-structure}). The calculation starts with the Dirac equation in the `Newtonian' gauge (Painlev\'e-Gullstrand coordinates) \cite{Lasenby:2002mc}, which can be written in Hamiltonian form $i \partial_t \psi = \hat{H} \psi$ where
\beq
\hat{H} \psi =  - i \gam^0 \gam^j \partial_j \psi + \mu \gam^0 \psi + \hat{H}_I \psi, \quad \quad \hat{H}_I \psi = i \sqrt{\frac{2M}{r}} \left( \frac{\partial}{\partial r} + \frac{3}{4r} \right) \psi, 
\eeq
with gamma matrices from the Dirac-Pauli representation. 
This resembles a flat-space equation with a novel interaction term. Applying a Foldy-Wouthuysen transformation \cite{Foldy:1950} leads to
\beq
\hat{H} = \gam^0 \mu + \hat{H}_0 + \hat{H}_1 + \ldots ,
\eeq
where 
\beq
\hat{H}_0 = \frac{1}{2\mu} \gam^0 \opO^2 + \opE ,
\quad \quad
\hat{H}_1 = -\frac{1}{8\mu^3} \gam^0 \opO^4 - \frac{1}{8 \mu^2} \left[ \opO, \left[ \opO, \opE \right] \right] ,
\label{eq:H1}
\eeq
and the Pauli-even and odd operators are
$\opO = -i \gam^0 \gam^i \partial_i$ 
and
$\opE = \hat{H}_I$. The Foldy-Wouthuysen transformation renders the Dirac equation in block-diagonal form, with upper (lower) components representing particles (anti-particles).

The zeroth-order equation $i \partial_t \psi =  \mu \gam^0 \psi + \hat{H}_0 \psi$ can be written in more familiar form by introducing the phase-transformed wavefunction, 
\beq
\psi =  e^{i M \mu \sqrt{8 r / M}} \begin{pmatrix} \Psi \\ 0 \end{pmatrix}, 
\eeq 
leading to
\begin{align}
(i \partial_t - \mu) \Psi &= -\frac{1}{2\mu} \bnab^2 \Psi - \frac{M\mu}{r} \Psi .  
\end{align}
Introducing a separation of variables, 
\beq
\Psi \equiv e^{-i \mu (1 + \mathcal{E}^{(0)}_{n} ) t} \Psi_{n\ell}, \quad \quad \Psi_{n\ell} = R_{n\ell}(r) Y_{lm}(\theta, \phi) \chi ,
\eeq 
where $\chi$ is any constant two-spinor, leads to a time-independent 1D Schrodinger equation with a $1/r$ `Newtonian' potential,
\begin{align}
\mathcal{E}^{(0)}_{n}  R_{n\ell} &= -\frac{1}{2\mu} \left( \frac{d^2}{dr^2} + \frac{2}{r} \frac{d}{dr} - \frac{\ell (\ell + 1)}{r^2} \right) R_{n\ell} - \frac{M\mu}{r} R_{n\ell} . 
\end{align}
This equation has standard hydrogenic solutions for the bound state energy levels $\mathcal{E}^{(0)}_{n} = -(M\mu)^2 / 2 n^2$ and wavefunctions,
\beq
R_{nl}(r) = A_{n\ell} e^{-x/2} x^\ell L_{n-\ell-1}^{2\ell + 1}(x) , 
\eeq
where $x = (2/n) (M\mu)^2 (r/M)$, $L_{\hat{n}}^{2\ell + 1}$ is an associated Laguerre polynomial, $Y_{lm}$ is a spherical harmonic, and $A_{n\ell}$ is a normalization constant.

To deduce the fine-structure correction to the energy, $\mathcal{E}_{n \ell j}^{(1)}$, one calculates the expectation values of terms in the first-order Hamiltonian $\hat{H}_1$, Eq.~(\ref{eq:H1}), when closed with the zeroth order wavefunctions, that is,
\beq
\mathcal{E}_{n \ell j}^{(1)} = \left< \Psi_{n\ell}^\ast e^{-i M \mu  \sqrt{8 r / M}} \right| \mu^{-1} \hat{H}_1 \left| e^{i M \mu  \sqrt{8 r / M}} \Psi_{n\ell}  \right> .
\eeq
The details of this calculation may be found on p100--105 of Ref.~\cite{Dolan:2007:thesis}. The fine-structure correction $\mathcal{E}_{n \ell j}^{(1)}$ is made up of two parts,
\beq
\left<  -\frac{1}{8\mu^4} \gam^0 \opO^4 \right> = -\frac{(M\mu)^4}{8n^4} \left(-15 + \frac{48n}{2\ell + 1} + \frac{9n}{\ell(\ell+1)(2\ell+1)} \right)  \label{eq:fs1}
\eeq
and
\beq
\left<  -\frac{1}{8\mu^3}  \left[ \opO, \left[ \opO, \opE \right] \right]  \right> = -\frac{(M\mu)^4}{8n^4} \frac{n (12 \lambda + 3)}{\ell(\ell+1)(2\ell+1)} . \label{eq:fs2}
\eeq
Here we have corrected an error in Ref.~\cite{Dolan:2007:thesis} in the coefficient of the second term of Eq.~(\ref{eq:fs1}) (this error may be traced back to the final entry of Table 5.1 of Ref.~\cite{Dolan:2007:thesis}, where the `4' should be a `2'). Taking the sum of contributions leads to Eq.~(\ref{eq:fine-structure}).


\bibliographystyle{iopart-num}
\bibliography{refs}

\providecommand{\newblock}{}
\begin{thebibliography}{10}
\expandafter\ifx\csname url\endcsname\relax
  \def\url#1{{\tt #1}}\fi
\expandafter\ifx\csname urlprefix\endcsname\relax\def\urlprefix{URL }\fi
\providecommand{\eprint}[2][]{\url{#2}}

\bibitem{Pauli:1933gc}
Pauli W 1933 {\em Annalen Phys.\/} {\bf 18S5} 337--372

\bibitem{Halpern:Heller:1935}
Halpern O and Heller G 1935 {\em Phys. Rev.\/} {\bf 48}(5) 434--438

\bibitem{Brill:1957fx}
Brill D~R and Wheeler J~A 1957 {\em Rev.Mod.Phys.\/} {\bf 29} 465--479

\bibitem{Nesvizhevsky:2002ef}
Nesvizhevsky V, Borner H, Petukhov A, Abele H, Baessler S {\em et~al.\/} 2002
  {\em Nature\/} {\bf 415} 297--299

\bibitem{Nesvizhevsky:2003ww}
Nesvizhevsky V, Borner H, Gagarski A, Petoukhov A, Petrov G {\em et~al.\/} 2003
  {\em Phys.Rev.\/} {\bf D67} 102002 (\textit{Preprint}
  \eprint{hep-ph/0306198})

\bibitem{Berti:2015itd}
Berti E, Barausse E, Cardoso V, Gualtieri L, Pani P {\em et~al.\/} 2015
  (\textit{Preprint} \eprint{1501.07274})

\bibitem{Arvanitaki:2009fg}
Arvanitaki A, Dimopoulos S, Dubovsky S, Kaloper N and March-Russell J 2010 {\em
  Phys.Rev.\/} {\bf D81} 123530 (\textit{Preprint} \eprint{0905.4720})

\bibitem{Arvanitaki:2014wva}
Arvanitaki A, Baryakhtar M and Huang X 2014  (\textit{Preprint}
  \eprint{1411.2263})

\bibitem{Nucamendi:2000jw}
Nucamendi U, Salgado M and Sudarsky D 2001 {\em Phys.Rev.\/} {\bf D63} 125016
  (\textit{Preprint} \eprint{gr-qc/0011049})

\bibitem{Marsh:2010wq}
Marsh D~J and Ferreira P~G 2010 {\em Phys.Rev.\/} {\bf D82} 103528
  (\textit{Preprint} \eprint{1009.3501})

\bibitem{Burt:2011pv}
Barranco J, Bernal A, Degollado J~C, Diez-Tejedor A, Megevand M {\em et~al.\/}
  2011 {\em Phys.Rev.\/} {\bf D84} 083008 (\textit{Preprint}
  \eprint{1108.0931})

\bibitem{Israel:1967za}
Israel W 1968 {\em Commun.Math.Phys.\/} {\bf 8} 245--260

\bibitem{Carter:1971zc}
Carter B 1971 {\em Phys.Rev.Lett.\/} {\bf 26} 331--333

\bibitem{Ruffini:1971bza}
Ruffini R and Wheeler J~A 1971 {\em Phys.Today\/} {\bf 24} 30

\bibitem{Bekenstein:1995un}
Bekenstein J 1995 {\em Phys.Rev.\/} {\bf D51} 6608--6611

\bibitem{Bekenstein:1996pn}
Bekenstein J~D 1996  (\textit{Preprint} \eprint{gr-qc/9605059})

\bibitem{Chrusciel:2012}
Chrusciel P~T, Costa J~L, Heusler M {\em et~al.\/} 2012 {\em Living Rev.
  Relativity\/} {\bf 15}

\bibitem{Press:1972zz}
Press W~H and Teukolsky S~A 1972 {\em Nature\/} {\bf 238} 211--212

\bibitem{Damour:1976kh}
Damour T, Deruelle N and Ruffini R 1976 {\em Lett.Nuovo Cim.\/} {\bf 15}
  257--262

\bibitem{Zouros:1979iw}
Zouros T and Eardley D 1979 {\em Annals Phys.\/} {\bf 118} 139--155

\bibitem{Furuhashi:2004jk}
Furuhashi H and Nambu Y 2004 {\em Prog.Theor.Phys.\/} {\bf 112} 983--995
  (\textit{Preprint} \eprint{gr-qc/0402037})

\bibitem{Cardoso:2004nk}
Cardoso V, Dias O~J, Lemos J~P and Yoshida S 2004 {\em Phys.Rev.\/} {\bf D70}
  044039 (\textit{Preprint} \eprint{hep-th/0404096})

\bibitem{Dolan:2007mj}
Dolan S~R 2007 {\em Phys.Rev.\/} {\bf D76} 084001 (\textit{Preprint}
  \eprint{0705.2880})

\bibitem{Dolan:2012yt}
Dolan S~R 2013 {\em Phys.Rev.\/} {\bf D87} 124026 (\textit{Preprint}
  \eprint{1212.1477})

\bibitem{Martellini:1977qf}
Martellini M and Treves A 1977 {\em Phys.Rev.\/} {\bf D15} 3060--3061

\bibitem{Iyer:1977cj}
Iyer B~R and Kumar A 1977 {\em Pramana\/} {\bf 8} 500--511

\bibitem{Iyer:1978du}
Iyer B~R and Kumar A 1978 {\em Phys.Rev.\/} {\bf D18} 4799--4801

\bibitem{Gueven:1977dq}
Gueven R 1977 {\em Phys.Rev.\/} {\bf D16} 1706--1711

\bibitem{Wagh:1986cz}
Wagh S and Dadhich N 1985 {\em Phys.Rev.\/} {\bf D32} 1863--1865

\bibitem{Brito:2015oca}
Brito R, Cardoso V and Pani P 2015  (\textit{Preprint} \eprint{1501.06570})

\bibitem{Brito:2014wla}
Brito R, Cardoso V and Pani P 2015 {\em Class.Quant.Grav.\/} {\bf 32} 134001
  (\textit{Preprint} \eprint{1411.0686})

\bibitem{Yoshino:2015nsa}
Yoshino H and Kodama H 2015  (\textit{Preprint} \eprint{1505.00714})

\bibitem{Zilhao:2015tya}
Zilh‹o M, Witek H and Cardoso V 2015  (\textit{Preprint} \eprint{1505.00797})

\bibitem{Detweiler:1980uk}
Detweiler S~L 1980 {\em Phys.Rev.\/} {\bf D22} 2323--2326

\bibitem{Rosa:2009ei}
Rosa J 2010 {\em JHEP\/} {\bf 1006} 015 (\textit{Preprint} \eprint{0912.1780})

\bibitem{Ternov:1978gq}
Ternov I, Khalilov V, Chizhov G and Gaina A~B 1978 {\em Sov.Phys.J.\/} {\bf 21}
  1200--1204

\bibitem{Ternov:1980st}
Ternov I, Gaina A~B and Chizhov G 1980 {\em Sov.Phys.J.\/} {\bf 23} 695--700

\bibitem{Gal'tsov:1983}
Gal'tsov D, Pomerantseva G and Chizhov G 1983 {\em Soviet Physics Journal\/}
  {\bf 26} 743--745 ISSN 0038-5697

\bibitem{Ternov:1986bf}
Ternov I, Gaina A~B and Chizhov G 1986 {\em Sov.J.Nucl.Phys.\/} {\bf 44} 343

\bibitem{Gaina:1988dp}
Gaina A and Ternov I 1986 {\em Sov.Astron.Lett.\/} {\bf 12} 394--396

\bibitem{Ternov:1988xy}
Ternov I and Gaina A~B 1988 {\em Sov.Phys.J.\/} {\bf 31} 157--163

\bibitem{Gaina:1988nf}
Gaina A~B and Ternov I 1988 {\em Sov.Phys.J.\/} {\bf 31} 830--834

\bibitem{Gaina:1989bf}
Gaina A~B and Ternov I 1989 {\em Vestn.Mosk.Univ.Fiz.Astron.\/} {\bf 30N2}
  22--28

\bibitem{Lasenby:2002mc}
Lasenby A, Doran C, Pritchard J, Caceres A and Dolan S 2005 {\em Phys.Rev.\/}
  {\bf D72} 105014 (\textit{Preprint} \eprint{gr-qc/0209090})

\bibitem{Sturm:2007yf}
Sturm I and Witte F 2007  (\textit{Preprint} \eprint{0707.2676})

\bibitem{Grain:2007gn}
Grain J and Barrau A 2008 {\em Eur.Phys.J.\/} {\bf C53} 641--648
  (\textit{Preprint} \eprint{hep-th/0701265})

\bibitem{Hartman:2009qu}
Hartman T, Song W and Strominger A 2009  (\textit{Preprint} \eprint{0912.4265})

\bibitem{Laptev:2006}
Laptev Y~P and Fil'Chenkov M 2006 {\em Astronomical and Astrophysical
  Transactions\/} {\bf 25} 33--42

\bibitem{Pekeris:1989}
Pekeris C and Frankowski K 1989 {\em Physical Review A\/} {\bf 39} 518

\bibitem{Gair:thesis}
Gair J 2002 {\em Generalised Tolman-Bondi cosmologies and Kerr metric atoms\/}
  Ph.D. thesis University of Cambridge

\bibitem{Dokuchaev:2014}
Dokuchaev V and Eroshenko Y~N 2014 {\em Advances in High Energy Physics\/} {\bf
  2014}

\bibitem{Gaina:1992}
Gaina A and Zaslavskii O 1992 {\em Classical and Quantum Gravity\/} {\bf 9} 667

\bibitem{Zhou:2013dra}
Zhou X~N, Du X~L, Yang K and Liu Y~X 2014 {\em Phys.Rev.\/} {\bf D89} 043006
  (\textit{Preprint} \eprint{1308.2863})

\bibitem{Barranco:2012qs}
Barranco J, Bernal A, Degollado J~C, Diez-Tejedor A, Megevand M {\em et~al.\/}
  2012 {\em Phys.Rev.Lett.\/} {\bf 109} 081102 (\textit{Preprint}
  \eprint{1207.2153})

\bibitem{Cho:2003qe}
Cho H~T 2003 {\em Phys.Rev.\/} {\bf D68} 024003 (\textit{Preprint}
  \eprint{gr-qc/0303078})

\bibitem{Jing:2005dt}
Jing J~l 2005 {\em Phys.Rev.\/} {\bf D71} 124006 (\textit{Preprint}
  \eprint{gr-qc/0502023})

\bibitem{Jing:2005pk}
Jing J~l and Pan Q~y 2005 {\em Nucl.Phys.\/} {\bf B728} 109--120
  (\textit{Preprint} \eprint{gr-qc/0506098})

\bibitem{Rosa:2011my}
Rosa J~G and Dolan S~R 2012 {\em Phys.Rev.\/} {\bf D85} 044043
  (\textit{Preprint} \eprint{1110.4494})

\bibitem{Witek:2012tr}
Witek H, Cardoso V, Ishibashi A and Sperhake U 2013 {\em Phys.Rev.\/} {\bf D87}
  043513 (\textit{Preprint} \eprint{1212.0551})

\bibitem{Pani:2012vp}
Pani P, Cardoso V, Gualtieri L, Berti E and Ishibashi A 2012 {\em
  Phys.Rev.Lett.\/} {\bf 109} 131102 (\textit{Preprint} \eprint{1209.0465})

\bibitem{Pani:2012bp}
Pani P, Cardoso V, Gualtieri L, Berti E and Ishibashi A 2012 {\em Phys.Rev.\/}
  {\bf D86} 104017 (\textit{Preprint} \eprint{1209.0773})

\bibitem{Unruh:1973}
Unruh W 1973 {\em Phys. Rev. Lett.\/} {\bf 31}(20) 1265--1267

\bibitem{Chandrasekhar:1976ap}
Chandrasekhar S 1976 {\em Proc.Roy.Soc.Lond.\/} {\bf A349} 571--575

\bibitem{Chandrasekhar:1983}
Chandrasekhar S 1983 {\em {The mathematical theory of black holes}\/} (Oxford
  University Press)

\bibitem{Carter:1979fe}
Carter B and Mclenaghan R 1979 {\em Phys.Rev.\/} {\bf D19} 1093--1097

\bibitem{Kamran:1984mb}
Kamran N and Mclenaghan R 1984 {\em J.Math.Phys.\/} {\bf 25} 1019--1027

\bibitem{McKellar:1993ej}
McKellar B, Thomson M~J and Stephenson G 1993 {\em J.Phys.\/} {\bf A26}
  3649--3657

\bibitem{Finster:1999ry}
Finster F, Kamran N, Smoller J and Yau S~T 2000 {\em Commun.Pure Appl.Math.\/}
  {\bf 53} 902--929 (\textit{Preprint} \eprint{gr-qc/9905047})

\bibitem{Finster:2000jz}
Finster F, Kamran N, Smoller J and Yau S~T 2003 {\em Adv.Theor.Math.Phys.\/}
  {\bf 7} 25--52 (\textit{Preprint} \eprint{gr-qc/0005088})

\bibitem{Finster:2001vn}
Finster F, Kamran N, Smoller J and Yau S~T 2002 {\em Commun.Math.Phys.\/} {\bf
  230} 201--244 (\textit{Preprint} \eprint{gr-qc/0107094})

\bibitem{Finster:2008bg}
Finster F, Kamran N, Smoller J and Yau S~T 2008  (\textit{Preprint}
  \eprint{0801.1423})

\bibitem{Belgiorno:2008hk}
Belgiorno F and Cacciatori S~L 2010 {\em J.Math.Phys.\/} {\bf 51} 033517
  (\textit{Preprint} \eprint{0803.2496})

\bibitem{Belgiorno:2008xn}
Belgiorno F and Cacciatori S~L 2009 {\em J.Phys.\/} {\bf A42} 135207
  (\textit{Preprint} \eprint{0807.4310})

\bibitem{Carter:1968ks}
Carter B 1968 {\em Commun.Math.Phys.\/} {\bf 10} 280

\bibitem{Wiltshire:2009zza}
Wiltshire D~L, Visser M and Scott S~M 2009 {\em {The Kerr spacetime: Rotating
  black holes in general relativity}\/} (Cambridge University Press)

\bibitem{poisson2004relativist}
Poisson E 2004 {\em A relativist's toolkit: the mathematics of black-hole
  mechanics\/} (Cambridge University Press)

\bibitem{Casals:2012es}
Casals M, Dolan S~R, Nolan B~C, Ottewill A~C and Winstanley E 2013 {\em
  Phys.Rev.\/} {\bf D87} 064027 (\textit{Preprint} \eprint{1207.7089})

\bibitem{Dolan:2009kj}
Dolan S and Gair J 2009 {\em Class.Quant.Grav.\/} {\bf 26} 175020
  (\textit{Preprint} \eprint{0905.2974})

\bibitem{Unruh:1974bw}
Unruh W 1974 {\em Phys.Rev.\/} {\bf D10} 3194--3205

\bibitem{Leaver:1985ax}
Leaver E 1985 {\em Proc.Roy.Soc.Lond.\/} {\bf A402} 285--298

\bibitem{Suffern}
Suffern K~G, Fackerell E~D and Cosgrove C~M 1983 {\em Journal of Mathematical
  Physics\/} {\bf 24} 1350--1358

\bibitem{Kalnins:1992bf}
Kalnins E and Miller W 1992 {\em J.Math.Phys.\/} {\bf 33} 286--296

\bibitem{Simmendinger:1999}
Simmendinger C, Wunderlin A and Pelster A 1999 {\em Phys. Rev. E\/} {\bf 59}(5)
  5344--5353

\bibitem{Yoshino:2012kn}
Yoshino H and Kodama H 2012 {\em Prog.Theor.Phys.\/} {\bf 128} 153--190
  (\textit{Preprint} \eprint{1203.5070})

\bibitem{Yoshino:2013ofa}
Yoshino H and Kodama H 2014 {\em PTEP\/} {\bf 2014} 043E02 (\textit{Preprint}
  \eprint{1312.2326})

\bibitem{Yoshino:2014wwa}
Yoshino H and Kodama H 2014  (\textit{Preprint} \eprint{1407.2030})

\bibitem{Barranco:2013rua}
Barranco J, Bernal A, Degollado J~C, Diez-Tejedor A, Megevand M {\em et~al.\/}
  2014 {\em Phys.Rev.\/} {\bf D89} 083006 (\textit{Preprint}
  \eprint{1312.5808})

\bibitem{Dolan:2007:thesis}
Dolan S~R 2007 {\em Scattering, absorption and emission by black holes\/} Ph.D.
  thesis University of Cambridge

\bibitem{Griffiths:1995}
Griffiths D~J and Harris E~G 1995 {\em Introduction to quantum mechanics\/}
  vol~2 (Prentice Hall New Jersey)

\bibitem{Page:1976jj}
Page D~N 1976 {\em Phys.Rev.\/} {\bf D14} 1509--1510

\bibitem{Hod:2012px}
Hod S 2012 {\em Phys.Rev.\/} {\bf D86} 104026 (\textit{Preprint}
  \eprint{1211.3202})

\bibitem{Hod:2013zza}
Hod S 2013 {\em Eur.Phys.J.\/} {\bf C73} 2378 (\textit{Preprint}
  \eprint{1311.5298})

\bibitem{Herdeiro:2014goa}
Herdeiro C~A~R and Radu E 2014 {\em Phys.Rev.Lett.\/} {\bf 112} 221101
  (\textit{Preprint} \eprint{1403.2757})

\bibitem{Benone:2014ssa}
Benone C~L, Crispino L~C, Herdeiro C and Radu E 2014 {\em Phys.Rev.\/} {\bf
  D90} 104024 (\textit{Preprint} \eprint{1409.1593})

\bibitem{Herdeiro:2015gia}
Herdeiro C and Radu E 2015  (\textit{Preprint} \eprint{1501.04319})

\bibitem{Starobinskii:1973}
Starobinskii A 1973 {\em Zh. Eksp. Teor. Fiz\/} {\bf 64} 48

\bibitem{Foldy:1950}
Foldy L~L and Wouthuysen S~A 1950 {\em Physical Review\/} {\bf 78} 29

\end{thebibliography}

\end{document}